\documentclass[sigconf]{acmart}
\AtBeginDocument{
  
}
\usepackage{xcolor}
\definecolor{mred}{rgb}{0.75, 0, 0}
\definecolor{mgreen}{rgb}{0,0.65,0}
\definecolor{mblue}{rgb}{0,0,.85}
\usepackage[utf8]{inputenc}
\usepackage{alphabeta}
\usepackage{lipsum}
\copyrightyear{2024}

\title{
Advancing {\color{mgreen} Technology} for {\color{mred} Humanity} and {\color{mblue} Earth}
(+{\color{mblue} Water}+{\color{mblue} Air})
}

\author{Steve Mann, Martin Cooper, Bran Ferren, Thomas M. Coughlin,
Paul Travers
}
\orcid{0000-0003-0363-3690}

\renewcommand{\shortauthors}{Mann, Cooper, Ferren, Coughlin, Travers}

\begin{abstract}
As technology advances, the integration of physical, virtual, and social worlds has led to a complex landscape of ``Realities'' such as Virtual Reality (VR), Augmented Reality (AR), metaverse, spatial computing, and other emerging paradigms.  This paper builds upon and refines the concept of eXtended Reality (XR) as the unifying framework that not only interpolates across these diverse realities but also
extrapolates (extends) to create entirely new possibilities.  XR is the ``physical spatial metaverse,'' bridging the physical world, the virtual world of artificial intelligence, and the social world of human interaction.
These three worlds define the Socio-Cyber-Physical Taxonomy of XR that allows us to identify underexplored research areas such as Diminished Reality (DR), and chart future directions to {\bf advance technology for people and planet}.
We highlight the six core properties of XR
for applications in sustainability, healthcare, frontline work, and daily life.  Central to this vision is the development of AI-driven wearable
technologies, such as the smart eyeglass, that sustainably extend human capabilities.
\end{abstract}


\begin{document}

\begin{teaserfigure}
  \includegraphics[width=\textwidth]{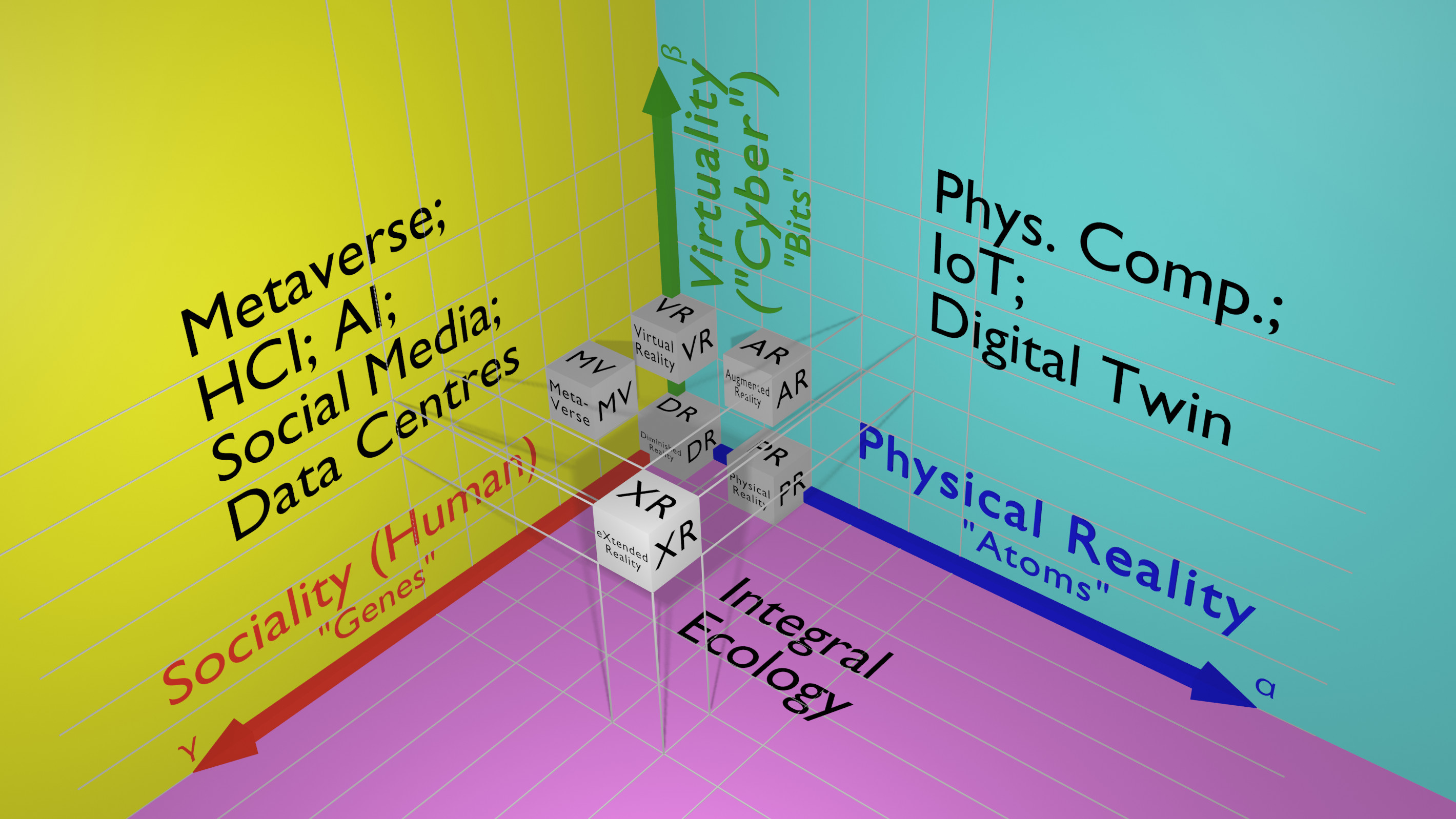}
  \Description[3D taxonomy of AR, VR, PR, etc., along axes of bits, atoms, and genes.]{A 3-dimensional taxonomy of the Realities, along the 3 axes of Physicality (Atoms), Virtuality (Bits), and Sociality (Genes) in which the Realities are shown as children's letter-blocks located with DR at the origin, PR along the Physicality axis, VR along the Virtuality axis, etc..}
  \caption{The Socio-Cyber-Physical Taxonomy:  Virtual Reality (VR) exists along the Virtuality (Cyber) axis (``Bits'').  Physical Reality (PR),  exists along the Physical Reality axis (``Atoms'').  AR (Augmented Reality) combines the world of ``Atoms'' and the world of ``Bits'' and exists in the Reality+Virtuality plane of Physical Computing and Internet of Things (IoT).  XR (eXtended Reality) spans the entire Socio-Cyber-Physical Space, interpolating between the various Realities and extrapolating beyond them~\protect\cite{mannwyckoff91, mann2001can, foster2024virtual, simge2024ingilizce, boffi2024co, hoffmann2024verschmelzen}.
  }
  \label{fig:teaser}
\end{teaserfigure}



\maketitle

\section{XR and (Im)mersivity}
Traditionally computers were large and cumbersome and were mainly associated with large indoor installations.  Eyewear such as Virtual Reality (VR) systems were also large and cumbersome, confined to labs, living rooms, or other indoor locations, where the user is isolated from the natural physical world.  As we bring these technologies into the real world, we have
Augmented Reality (AR) overlays combining the real world and the virtual world (Fig.~\ref{fig:teaser}).

XR (eXtended Reality) brings together the physical world, virtual world, and social world (Fig.~\ref{fig:teaser}) to connect us to each other and our surroundings.  XR is technology for humanity and Earth.  Tech for people and planet!

We begin with a brief look at the Socio-Cyber-Physical Space upon which XR is founded (Fig.~\ref{fig:teaser}), then motivate it through the example and use-case of wearable AI, and then revisit the concept of
Socio-Cyber-Physical space in detail in Section~\ref{sec:XRspace}.

\subsection{XR is social, technological, and physical}
XR is, by its very nature, technology that connects us to each other and to the physical world around us, as outlined in the Socio-Cyber-Physical Space of Fig.~\ref{fig:teaser} which forms a convenient taxonomy/ontology of the `R's (``Realities'').  VR (Virtual Reality) exists entirely along the Virtuality axis.  PR (Physical Reality), i.e. the real world, exists along the Reality axis of the physical world made of atoms (earth, water, air, etc.).  AR (Augmented Reality) combines the world of ``atoms'' and the world of ``bits'' and exists in the Reality+Virtuality plane.
Technologies that attenuate or diminish our senses (e.g. ear plugs, dark sunglasses, welding helmets, baseball caps) are examples of DR (Diminished Reality) which exists near the origin of the space (where the axes meet).  HCI (Human-Computer Interaction) and the Metaverse (shared social VR) exist in the Virtuality+Sociality plane.  XR (eXtended Reality) spans the entire Socio-Cyber-Physical space, interpolating between the various Realities and extrapolating beyond them~\protect\cite{mannwyckoff91, mann2001can, foster2024virtual, simge2024ingilizce, boffi2024co, hoffmann2024verschmelzen}.
Fig.~\ref{fig:teaser} is a simplified version of much of the work outlined in a previous IEEE Special Issue (see the introductory text~\protect\cite{el2024metaverse}), and associated preprint~\cite{mann2023extended}.

\section{Wearable AI: Introduction and motivation}
A good example of XR is Wearable AI~\cite{wearableai, mannwyckoff91} (Wearable Artificial Intelligence) which can take the form of a smart eyeglass.
The combination of XR and AI is also known as XI (eXtended Intelligence)~\cite{mannwyckoff91, cxi2019ieee}.

Wearable AI~\cite{wearableai} in the form of an
AI-Driven eyeglass
(``A-Eyeglass'')
offers capabilities that no other form factor can support.  
For example, XR takes the form of contextual information displayed directly upon the user’s field of view.  
AI insights using machine learning algorithms provide predictive analysis and recommendations. Hands-free operation now enables seamless multitasking, especially in critical or high-stress environments.  
Integration with Internet of Things (IoT) and other cyber-physical systems (Fig.~\ref{fig:teaser}) enables real-time updates and remote access.  
Moreover XR allows us to see beyond mere overlays, e.g. so that we can see in the ultravoilet, infrared, electromagnetic, and acoustic spectrum and share these visions with others since XR is shared and collaborative by its very nature~\cite{mannws, mannwyckoff91}.

The AI-driven smart eyeglass represent a pivotal advancement in wearable technology, as it embodies XR by combining Augmented Reality (AR), Artificial Intelligence (AI), connectivity, Human-Computer Interaction, Metaverse, and Internet of Things, to extend human interaction within the physical, virtual, and social worlds. These devices provide tailored solutions for sustainable healthcare, support for frontline workers, and optimized day-to-day activities such as sports and personal health. By focusing on sustainability, the smart eyeglass aligns with global efforts to improve health outcomes, reduce resource consumption, and foster a connected, data-driven lifestyle. All this while allowing people to get their heads out of their phones and back into the real world with the rest of humanity and earth.

A smart eyeglass powered by AI can extend human capabilities (XI = eXtended Intelligence) by integrating real-time data processing, contextual insights, and advanced user interfaces into wearable technology. The ability to extend physical reality with computational intelligence and social intelligence positions these devices as indispensable tools across industries and everyday life. Wearable AI such as embodied by the smart eyeglass is one very effective way to create a link connecting the virtual (artificial), real (physical), and social worlds especially as they are driven by AI. 
In a world increasingly focused on sustainability, the smart eyeglass offers scalable efficient solutions for healthcare, frontline workers, and individual users.
Healthcare is at the forefront of leveraging the AI-eyeglass for sustainable and impactful solutions.
For the patient directly, remote patient monitoring using an AI driven eyeglass enables healthcare professionals to monitor patients remotely through XR-based dashboards, reducing travel needs and resource consumption.  Sensors in the eyeglass measure what's happening around the patient (environment) as well as what's happening inside the patient's body (``invironment'') and, most importantly, the relationship between these (e.g. correlating electrocardiogram irregularities with environmental factors as captured by outward-facing cameras in the eyeglass).

AI algorithms analyze patient (invironment) data, together with surroundings (environment) providing early warnings for critical health events. This approach can minimize hospital admissions, cuts costs, and improve patient outcomes.
In operating rooms, the smart eyeglass is currently providing surgeons with XR-guided visuals of critical anatomy, AI-powered diagnostics, and procedural checklists, enhancing precision and reducing errors. The reduction of surgical errors also results in better resource utilization.
Medical students and professionals use the smart eyeglass for XR training from remote support to a first person perspective allowing students or experts to select between a bird's-eye-view or a ``Point-of-Eye'' view of live procedures. These applications reduce the need for physical resources like cadavers or expensive training setups.

Empowering frontline workers in industries such as construction, logistics, and emergency response are increasingly relying on the AI/XR-eyeglass for safety, efficien-cy, and sustainability.
AI algorithms integrated into an eyeglass can identify workplace hazards, such as structural instability or hazardous materials, alerting workers in real-time and preventing accidents.
Frontline workers can receive XR overlays and XR sensory extension with step-by-step instructions, reducing errors and training time. This is particularly useful in scenarios where manuals or printed materials are impractical.
A smart eyeglass equipped with environmental sensors and AI analytics supports emergency responders in assessing dangerous situations, tracking personnel, and coordinating rescue efforts more efficiently.

AI insights help optimize material usage, equipment operation, and energy consumption, promoting environmentally conscious practices in industries that rely on frontline labor.

{\bf The smart eyeglass is revolutionizing the way individuals approach personal health, fitness, and daily activities.}

Athletes use the smart eyeglass for real-time performance metrics, such as speed, distance, and biofeedback. AI-powered coaching delivers personalized insights, helping users optimize training while preventing injuries.
Casual fitness enthusiasts benefit from wearable AI~\cite{wearableai} that monitors activity levels, heart rate, calorie consumption, and environmental factors, providing actionable recommendations for a healthier lifestyle.

The smart eyeglass tracks vital signs such as oxygen levels, heart rate variability (HRV), and sleep patterns, helping users maintain optimal health. AI integration ensures early detection of anomalies, promoting preventative care.
For users in urban environments, the smart eyeglass facilitates navigation, communication, and extended reality overlays, together with extended senses, for improved situational awareness. Whether finding a bike path or checking air quality, or seeing in complete darkness, or seeing a safeyfield (path of safest travel), the AI-eyelgass provides practical solutions to everyday challenges.

The adoption of the AI-eyeglass is not just about convenience, it also aligns with broader sustainability goals, by enabling remote healthcare, minimizing travel for remote support and work, and streamlining industrial processes.  The smart eyeglass contributes to significant reductions in carbon emissions.
For example, the ability to see in complete darkness reduces the need for electric lighting.

The smart eyeglass improves efficiency in material and energy use, particularly in industries like manufacturing and logistics, promoting a circular economy.
Affordable smart eyeglasses powered by AI can bring advanced capabilities to underserved populations, bridge the digital divide, and foster equitable access to technology.

{\bf Wearability and fashion-forward design are critical for the success of smart eyeglasses.}

However, to be successful a key performance metric for smart eyeglasses is wearability and fashion. For many years, manufacturers have only been offering bulky, heavy, power-hungry and strange-looking devices.  Only recently the technology is finally becoming available to deliver the lightweight all-day wearable fashion forward solutions that are critical to the industry's success.
Wearability and fashion are critical factors for the success of smart eyeglasses, as they directly influence user adoption, satisfaction, and sustained use. 
A Smart eyeglass must be comfortable, stylish, and versatile to succeed in the marketplace. Consumers view such a product as both a functional tool and a personal accessory, so design must reflect a careful balance of functionality, comfort, and aesthetics. 

{\bf AI eyeglasses are on the verge of completely changing the socio-cyber-physical relationship, connecting people, planet, and technology.}

AI-eyeglasses represent a transformative shift in how humans interact with technology and the world around them. From revolutionizing healthcare to empowering frontline workers and enhancing daily life, these devices foster a profound connection between people, technology, and the environment. By focusing on sustainable and scalable applications, the smart eyeglass stands as a beacon of hope for a future where technology serves both humanity and the planet.

XR in the form of wearable AI-glass will:
\begin{itemize}
\item enhance sustainability through reduced resource use and improved efficiency.
\item have applications in healthcare and frontline work and showcase their transformative potential;
\item work for
everyday use cases that bridge the gap between advanced technology and accessible, meaningful benefits for individuals;
\item not only deliver innovative features but also integrate seamlessly into users' lives and wardrobes.
\end{itemize}

We envision a future where smart glasses not only improve lives but also contribute to a healthier planet.

\section{Historical and philosophical background}
To fully appreciate the benefit of XR above and beyond AR, we need to look at the historical background leading up to XR and Metavision/Metaveillance (and metaverse).

\subsection{Technology that truly connects us to each other and our surroundings}
While much has been done with social media and collaborative computing in an effort to connect us, an important goal of XR has always been to connect us in a profound way.  This connection can (and often should) go beyond merely collaborating in the virtual world, to include the social and physical worlds.  Consider some historical examples of XR.

Firstly, consider the S.W.I.M., an early spatial computing system developed in Canada in the early 1970s~\cite{impulse, soundfieldshirt, campuscanada, mannws, kineveillance, mann2020ichms, mann2018phenomenological}.  This system comprised arrays of computer-controlled light sources set in motion in a darkened room, and arranged in such a way as to allow viewers to see electromagnetic radio waves, sound waves, and other physical phenomena. SWIM is based on very broad-band hardware, so it updates in less than 100 nanoseconds, thus having an effective ``frame rate'' of many megahertz.

\begin{figure}
    \centering
    \includegraphics[width=\linewidth]{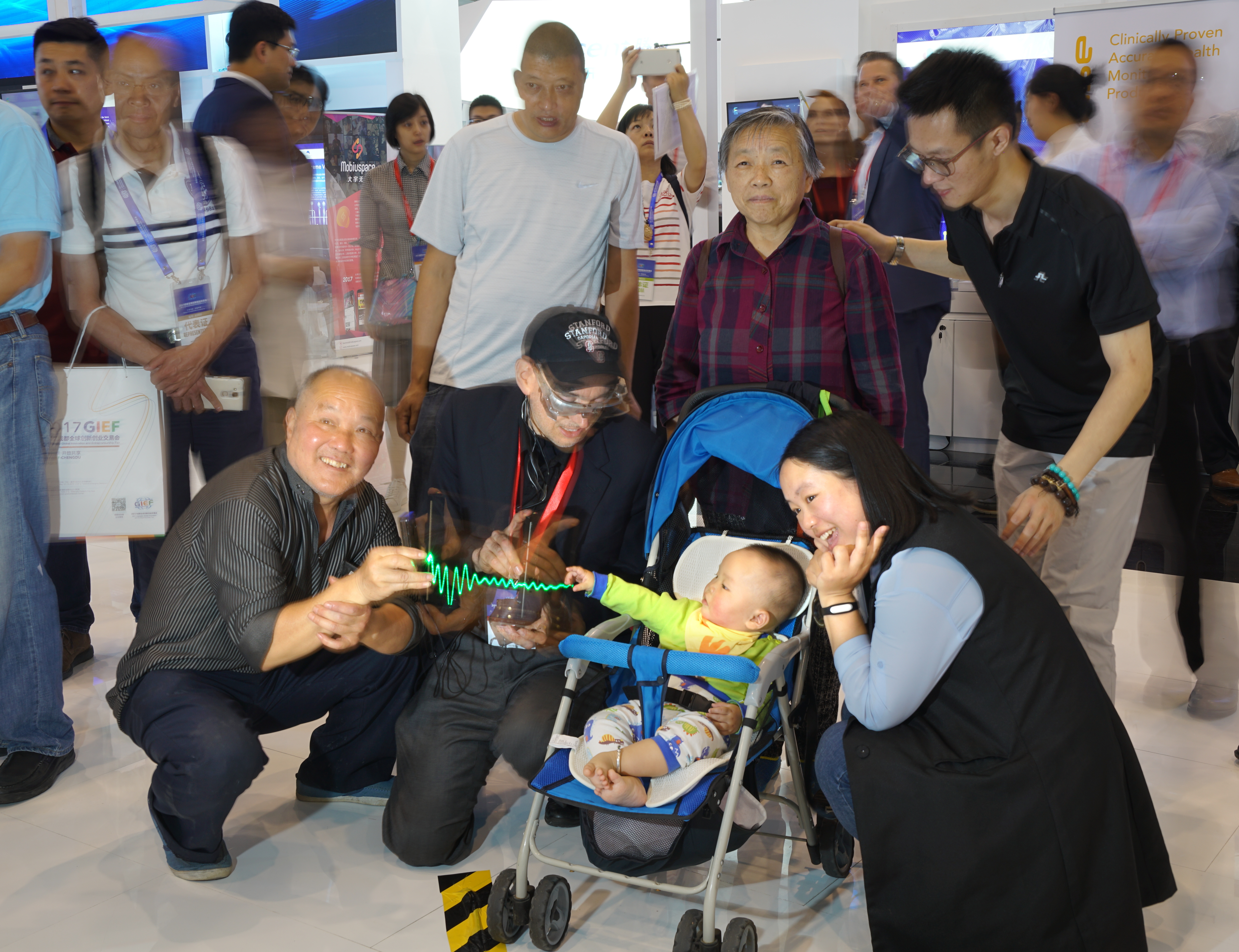}
    \Description[Photograph of SWIM at a tradeshow.]{Photograph of the SWIM (Sequential Wave Imprinting Machine) comprised of a wristworn array of 100 green LEDs (Light Emitting Diodes) waved back-and-forth to show a waveform in space.  Also shown is a child in a baby stroller interacting with the waveform amid a crowd of attendees at the tradeshow.}
    \caption{S.W.I.M. (Sequential Wave Imprinting Machine) is a spatial computing system invented in Canada in the early 1970s that allows large numbers of people to see, understand, grasp, touch, and feel electromagnetic radio waves, sound waves, and other phenomena at exact 1:1 scale with perfect alignment between the physical and virtual content, updated at effectively millions of frames per second (instantaneously)~\protect\cite{mannwyckoff91}.
    It may be used with the XR eyewear, but it can also be seen by thousands of people in the surrounding space without the need for any special eyewear, as shown to an extremely large audience at this recent tradeshow.}
    \label{fig:swim}
\end{figure}

What was unique about SWIM was that it facilitated direct human interaction among large groups (e.g. hundreds or thousands) of people, able to see and understand things happening beyond the normal range of human perception~\cite{mannwyckoff91}.
See Fig.~\ref{fig:swim}.

\begin{figure}
    \centering
    \includegraphics[height=.33\linewidth]{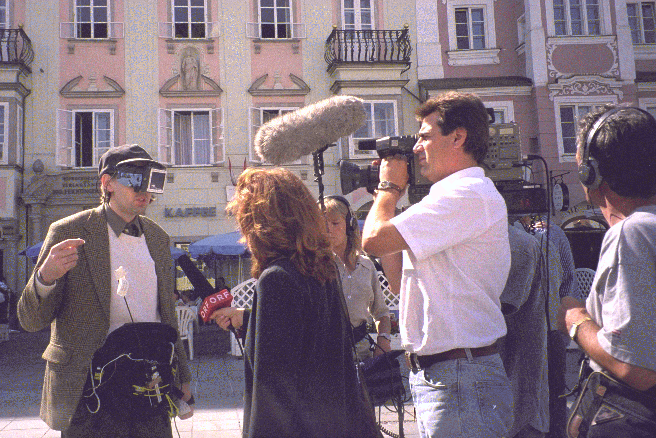}
    \includegraphics[height=.33\linewidth]{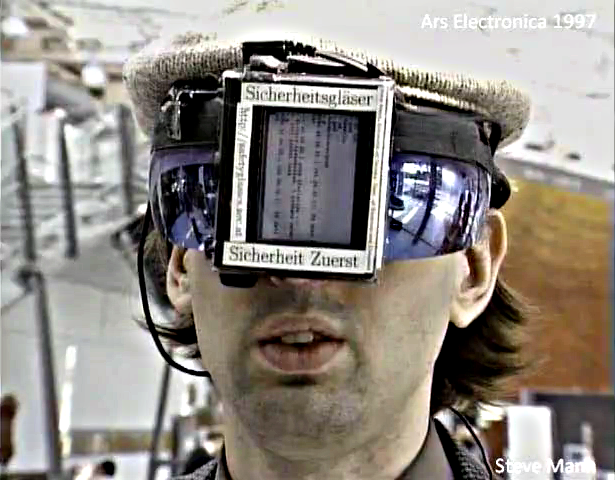}
    \Description[Photo of eyeglass with inwards and outwards-facing display.]{Photograph of an eyeglass having both inwards and outwards-facing displays, as shown at a public art performance, with a crowd of media representatives surrounding a man wearing the eyeglass.  Some of the media representatives have professional audio and video equipment such as large microphones and professional broadcast-quality video equipment.  The setting is a downtown urban area with buildings in the background.}
    \caption{Sicherheitsglaeser presented at Ars Electronica 1997 used a wearable computer with both an inwards-facing eyeglass display for the wearer to see, as well as an outwards-facing eyeglass display (on the same eyeglass) for local participants to see.  It facilitated a direct and meaningful connection between the virtual, physical, and social worlds of more than 30,000 remote participants and hundreds of local participants as a form of performance art.}
    \label{fig:sicherheitsglaeser}
\end{figure}

As a second example of XR, consider Sicherheitsglaeser, a live performance presented at Ars Electronica Sept. 1997,  based on concepts presented the conference~\cite{mannars}, followed by an art exhibit at List Visual Arts Centre Oct-Dec 1997~\cite{mannarsperformance}.  It comprised of a wearable computer supporting a large remote audience of more than 30,000 people, engaged in an XR (eXtended Reality) metavision experience, while at the same time featuring both an inwards-facing display for the wearer, and an outwards-facing display for a local audience.  In this way Sicherheitsglaeser facilitated direct interaction between remote and local participants.  See Fig.~\ref{fig:sicherheitsglaeser}.

\begin{figure*}
    \centering
    \includegraphics[height=2.56in]{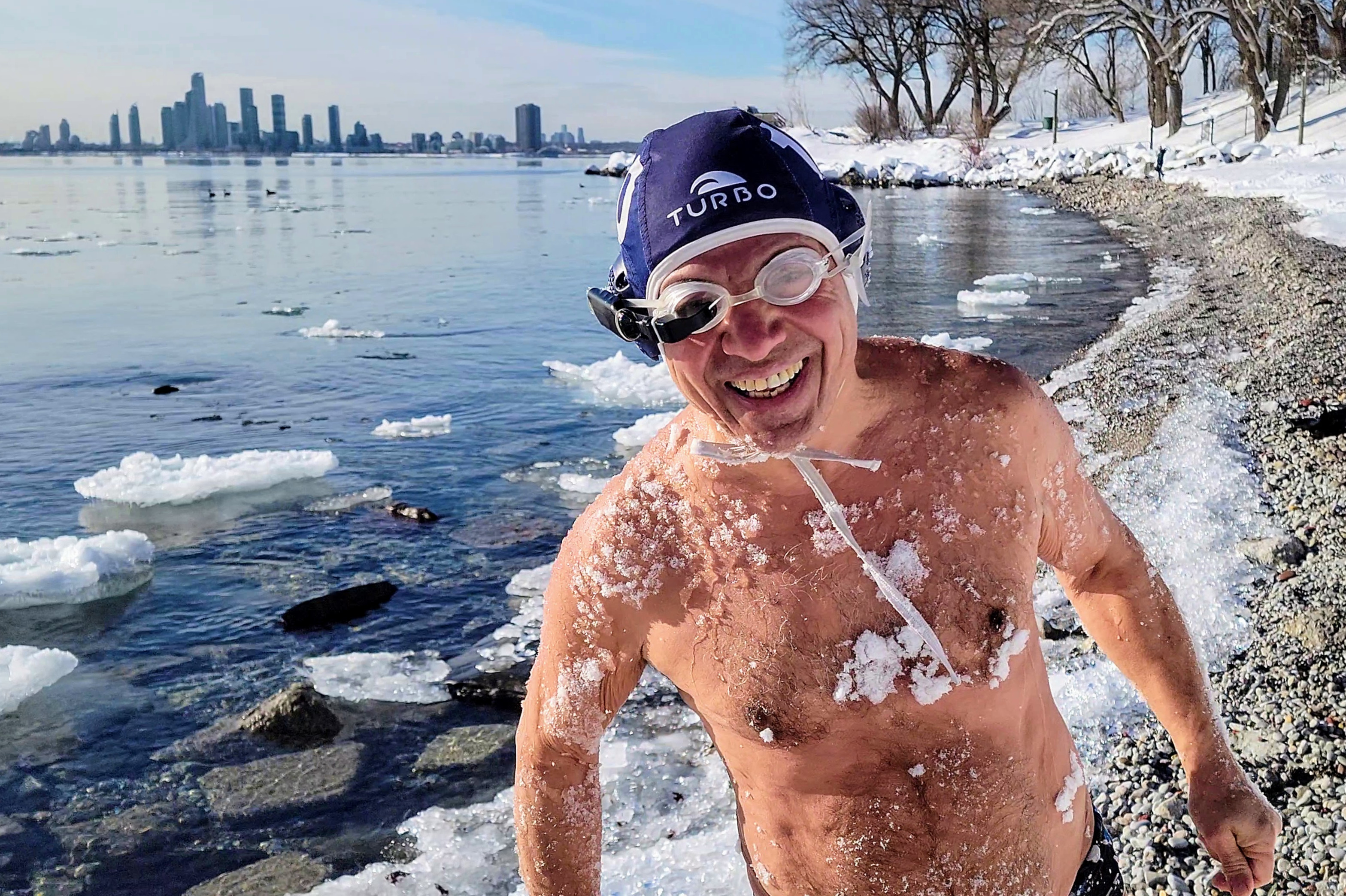}
    \includegraphics[height=2.56in]{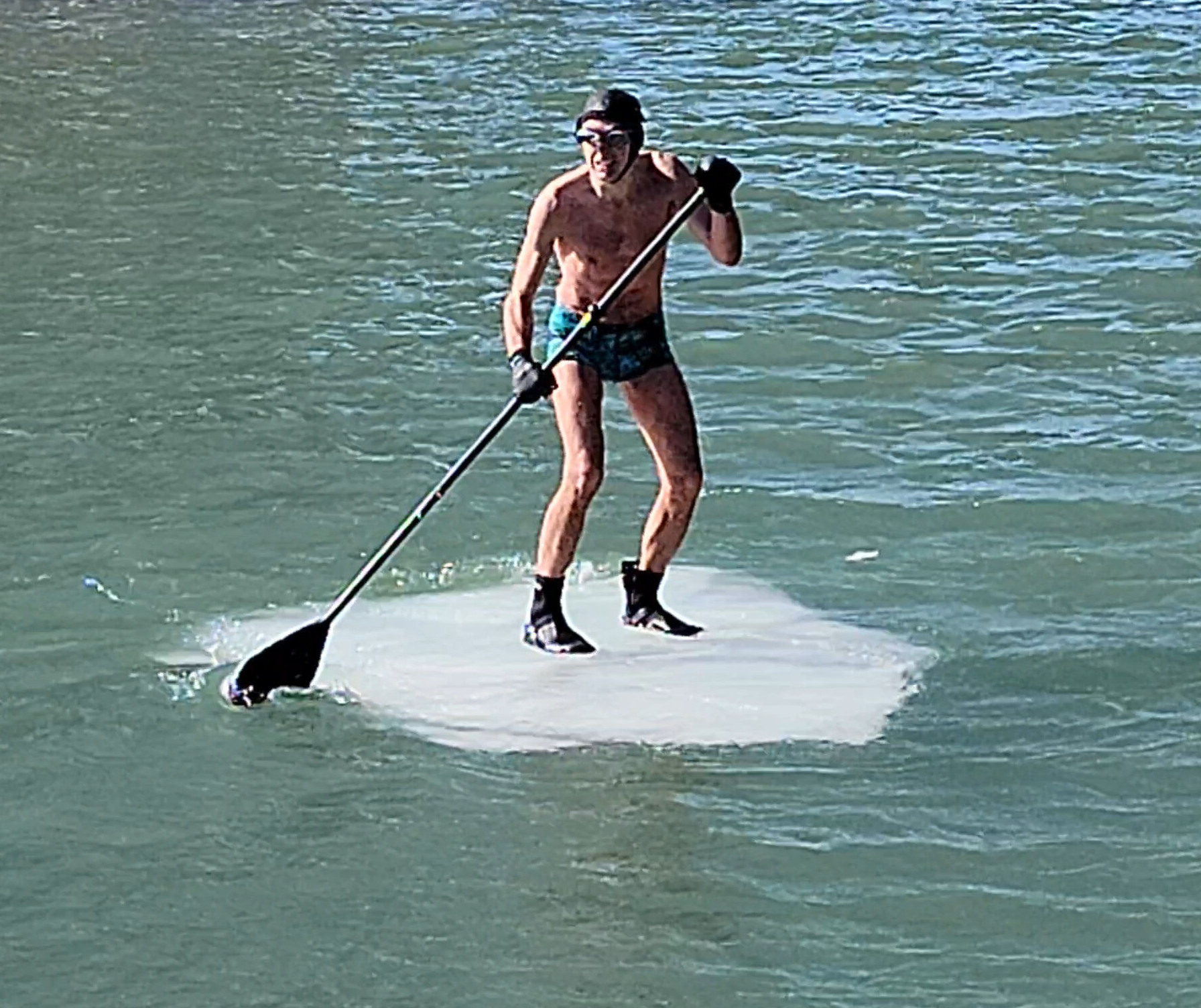}
    \includegraphics[width=\linewidth]{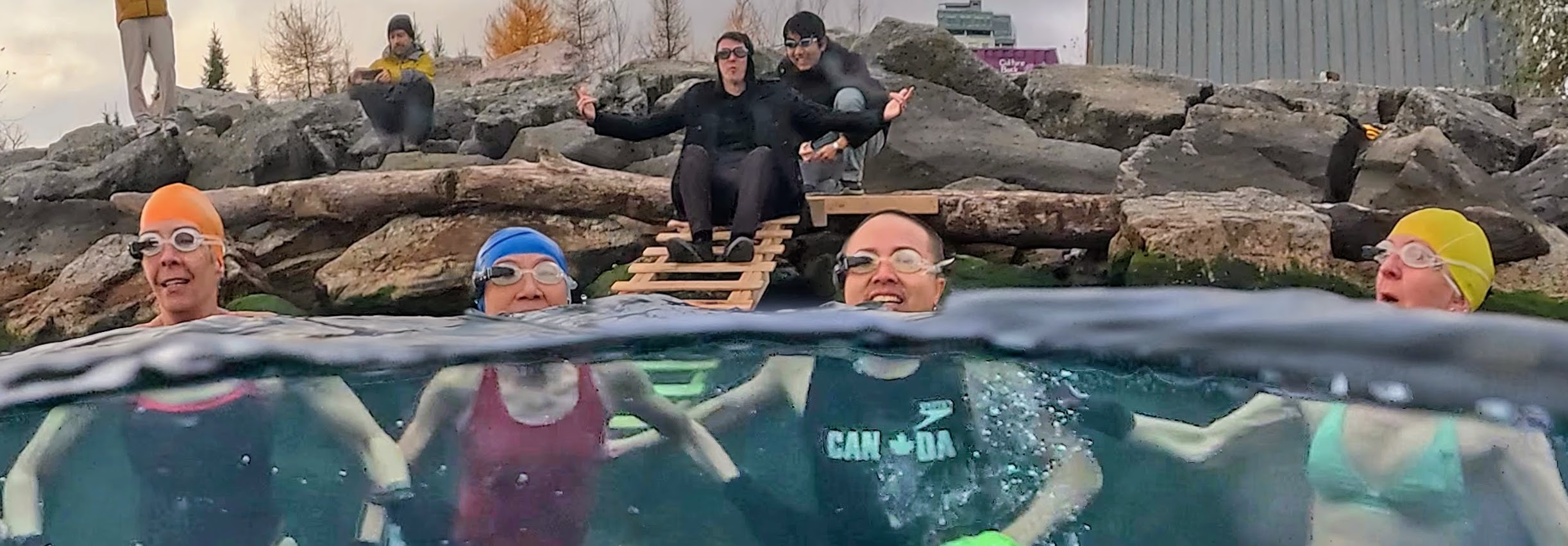}
    \Description[Underwater photos showing underwater XR glass.]{Photographs of icewater swimmers wearing XR glass.  One swimmer in swim briefs is covered in snow on a snow-covered beach with water and ice floes in the background.  In another picture a man in swim briefs stands on a piece of ice paddling.  In another picture there are 4 swimmers wearing XR glass and two people on a dock wearing XR glass as well.  The picture shows partly below water and above water.  The heads of the 4 swimmers are above water and their bodies are below water.  The dock is partly above and partly below water.  The portions of the image below and above water align properly  (matching of the refractive index of the water).}
    \caption{Example of XR for situational awareness during icewater swimming.  A wearable computer with XR eyeglass connects each participant to the others and to their surroundings.  XR is about connecting us to each other and our surroundings, without imposing limits on what we can do or experience.}
    \label{fig:smartswim}
\end{figure*}

As a third example of XR, consider the use of wearable computing during an icewater swim that allows a group of swimmers to remain in contact with each other and with remote members of the swim group for safety and situational awareness, as illustrated in Fig.~\ref{fig:smartswim}.  Wearable computers, such as Vuzix SmartSwim, allow the partipants to each their own, as well as each-others' vital signs, heart, respiration, and brain activity (e.g. blood-oxygen levels in the brain by way of fNIRS which is more reliable than EEG underwater).  Additionally swimmers see their own and each other's kolympographic~\cite{mann2021water} information including map, location, heading, direction, etc., as well as water temperature, wave-height data, and water quality data.

\begin{figure}
    \centering
    \includegraphics[width=\linewidth]{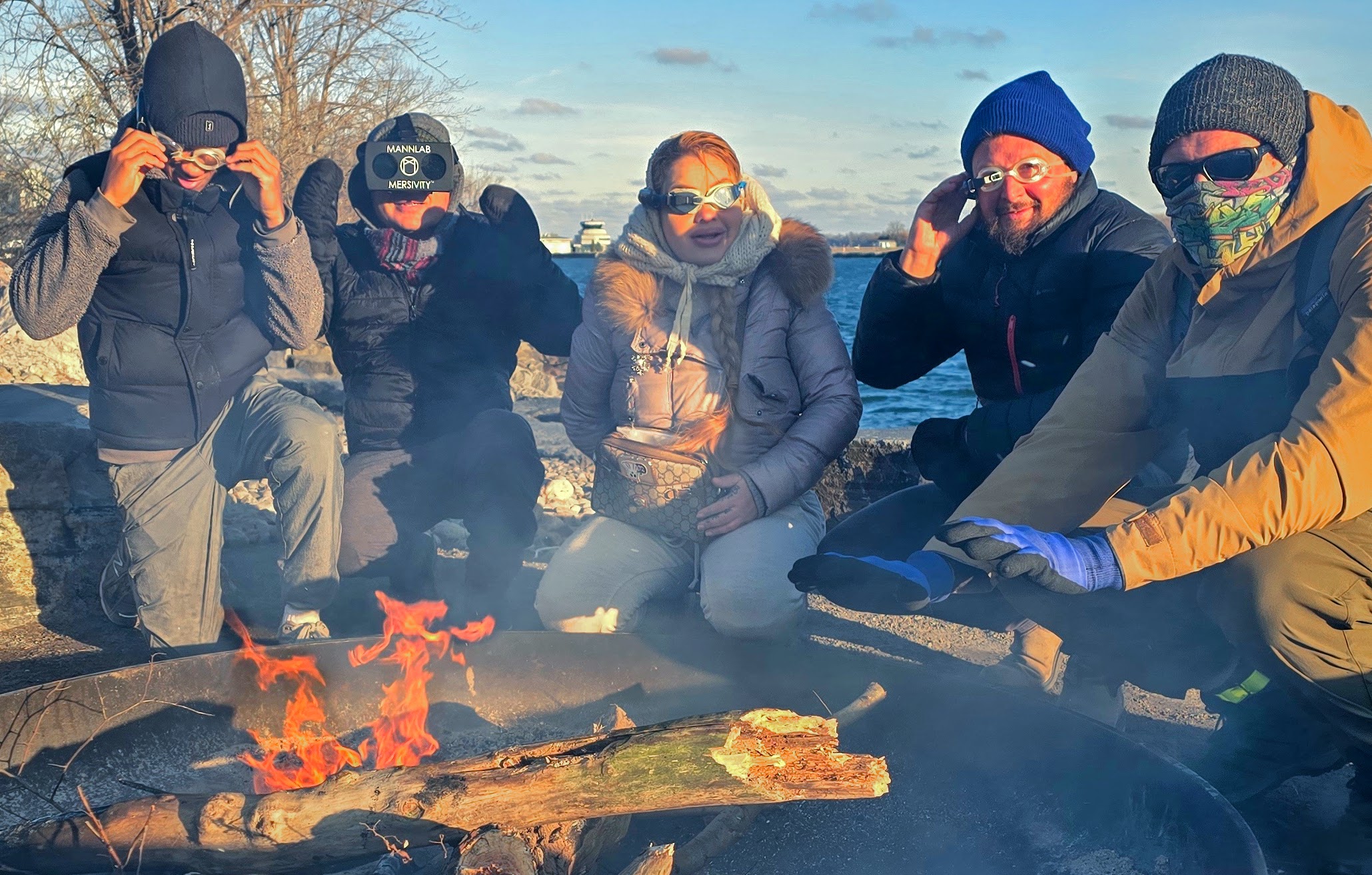}
    \Description[Photo of a group wearing VR/XR glass around a campfire.]{Photograph of 5 people wearing VR / XR glass setting in front of a circular steel firepit with a campfire burning in front of them.  One man warms his hands over the fire, while 2 others adjust their XR eyewear.}
    \caption{CampfireVR/XR: Technology should connect us to our environment, and should work everwhere, not just in the home or factory.  The MannLab Mersivity underwater VR glass as well as the Vuzix SmartSwim are examples of technology that works well in nearly any environment.}
    \label{fig:campfirevr}
\end{figure}

As a fourth example, consider the possibility for XR to connect us to each other and to the wilderness, nature, etc., such as at a campfire (Fig.~\ref{fig:campfirevr}).


It should be emphasized here that XR as a wearable immersive technology has the capacity to bring about a deeply profound connection between humans by connecting them to each other and their surroundings/environment/world.

We are all part of each others' surroundings, e.g. any one of us could truthfully say: ``You are part of my environment and I am part of your environment.''. Thus we cannot over-stress the importance of the full and complete Socio-Cyber-Physical connection afforded by XR.

\section{Mersivity}
The examples of XR provided in the previous section highlight
examples of Mersivity.  Mersivity regards 
technology as a generalized vessel that connects us to each other and to our surroundings.

\begin{figure*}
    \centering
    \includegraphics[width=\linewidth]{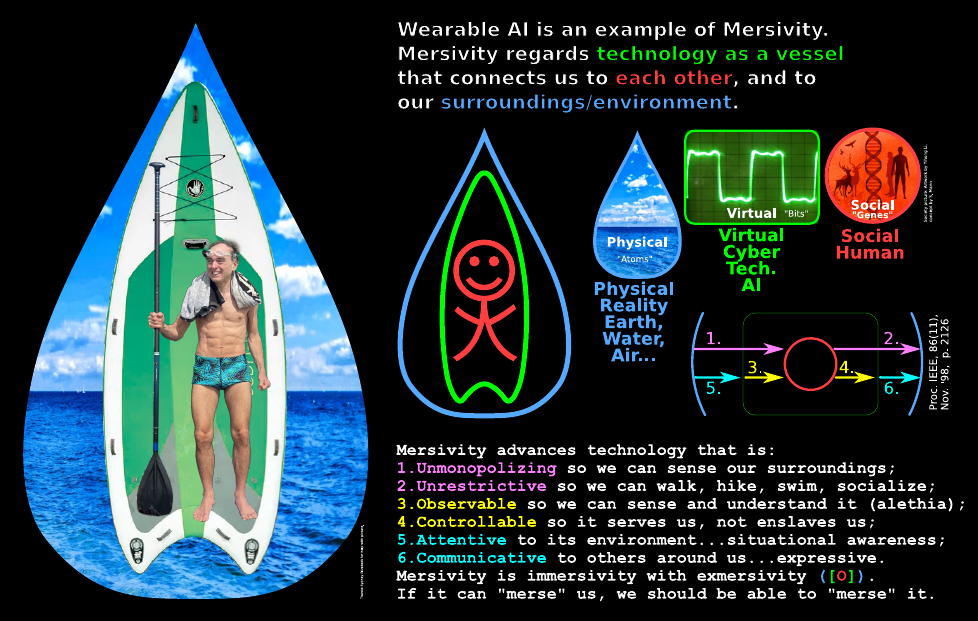}
    \Description[Mersivity diagram.]{Mersivity diagram with a person inside a vessel inside a Mochoid (similar to a water drop) shape.}
    \caption{Technology as a generalized vessel that has a symmetry of connectivity (i.e. both immersive and exmersive).  The resulting signal-flow symmetry to/from and through the technology gives us six signal flow paths and therefore six desirable properties for Mersivity.}
    \label{fig:vessel}
\end{figure*}

We can consider these connections more formally as follows.
Referring to Fig.~\ref{fig:vessel}

This set of connections define Mersivity, i.e. the way in which
we can regard technology as a generalized vessel (eyeglass, shoes, clothes, boat, car, building, etc.) denoted here as a smart paddleboard that connects us to each other and to our surroundings.  Such technologies are im{\bf mersive} and can encapsulate or enclose us to some degree.  We believe that technology should also be exmersive, e.g. durable enough as to not encumber us, We like technologies that can suit our lifestyle, e.g. allow us to go hiking in the wilderness, go for a swim, or paddle in the water. Othewrise we run the risk that technologies could imprison us and keep us away from our surroundings/environment. In this way technology must serve not only humanity but also nature/surroundings/environmnent/earth/water/air....

\section{The XR ({\color{mred} Socio}-{\color{mgreen} Cyber}-{\color{mblue}Physical}) Space \label{sec:XRspace}}
Wearable technologies that can merge, mix, mediate, modify, and remix the physical, virtual, and social worlds were created fifty years ago, in Canada, with the invention of Metavision and Metaveillance (1974) and SWIM (1974), an early form of wearable spatial computer~\cite{impulse, soundfieldshirt, campuscanada, mannws, kineveillance, mann2020ichms, mann2018phenomenological}.

VR (Virtual Reality), a term coined by 
Artaud in 1958~\cite{artaud}, and exemplified by 
the stereoscopic viewing apparatus of 
Sir Charles Wheatstone in 1838,
Morton Heilig's Sensorama, of 1956 (the first VR machine), and Ivan Sutherland's machine of 1968~\cite{sutherland68}, was characterized by an apparatus within a fixed setting, typically an indoor setting, while seated.

What sets wearable technologies apart from this tradition is the emphasis on connecting people to their surroundings, and to each other, e.g. being able to walk, run, swim, etc., while wearing and using the technology.  Along with these wearable technologies has come a dizzying and confusing collection of often conflicting terminology. Along with VR (Virtual Reality), we now have many other `R's such as: AR (Augmented Reality); MR (Mediated Reality); IR (Intelligent Reality), SR (Spatial Reality) or spatial computing; and Metaverse.

\begin{figure}
    \centering
    \includegraphics[width=\linewidth]{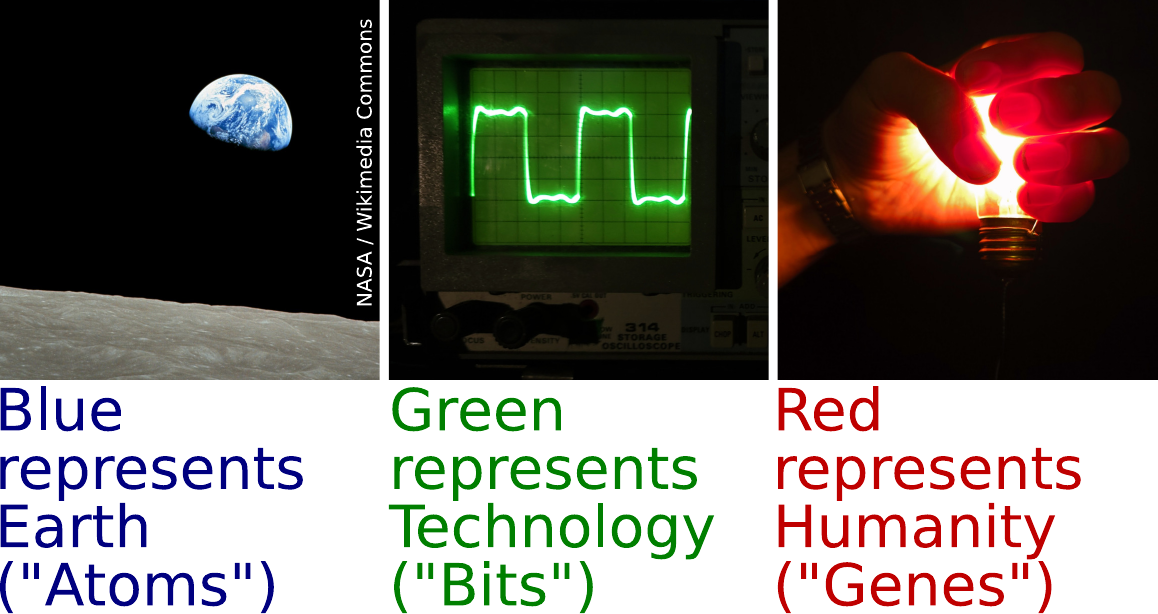}
    \Description[Three photos to represent each of the colors.]{Earthrise photograph on the left representing the blue planet, Oscilloscope photograph representing the green of technology, and a white light bulb held in the hand, showing the red glow coming through.}
    \caption{NASA Earthrise photo showing the blue planet Earth.  Photograph of an oscilloscope showing the green trace, representing ``Bits''.  Photograph of a human hand holding a white light bulb in the dark, showing  the red glow through the blood under the flesh.}
    \label{fig:earthriseTek314bulb}
\end{figure}

We seek to clarify this confusion through revisiting the concept of eXtended Reality (XR). We begin by discussing how the concept of the XRspace (Fig.~\ref{fig:teaser}) can be represented as a taxonomy or ontology of ``Realities'' along three axes: Physical, Virtual, and Social. For simplicity the axes are color-coded, using blue to represent the Physicality (``Earth'') axis, green to represent the Virtuality ("Cyber" / "Tech") axis, and red to represent the Humanity axis.  The color scheme for these three axes follows the illustration of Fig.~\ref{fig:earthriseTek314bulb}.

 


Let us begin with ``Atoms'' and ``Bits'', the cyan-colored (blue and green) Physical Computing plane of Fig.~\ref{fig:teaser}.

\subsection{Atoms and Bits}
Consider two axes: the physical world of ``atoms'' (earth, water, air, etc.) and the virtual/cyber world of ``bits''.
Together these axes illustrate ``Physical Computing'', i.e. a ``Cyber-Physical'' System (CPS), defined by Physicality (``atoms'') and Virtuality (``bits'') axes in Fig.~\ref{fig:teaser}.

This is a simplified version of the figure previously outlined in a preprint~\cite{mann2023extended} to a past IEEE Special Issue~\protect\cite{el2024metaverse}.
Together, the Physical axis (denoted in blue) and the Virtual axis (denoted in green) define the Cyber-Physical Plane (denoted in Fig.~\ref{fig:teaser} in cyan, which is a color that contains equal parts blue and green).

\begin{figure}
    \centering
    \includegraphics[width=\linewidth]{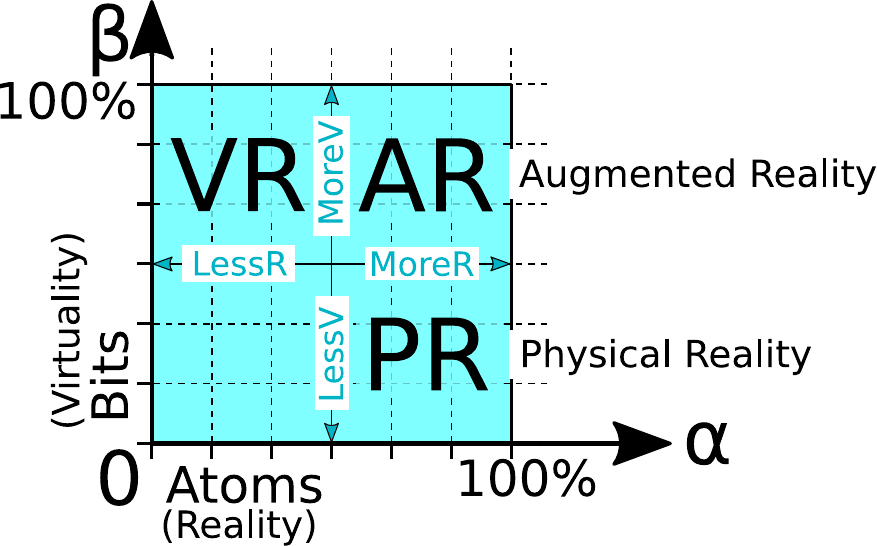}
    \Description[2D taxonomy of AR, VR, PR, and DR, along axes of bits and atoms.]{A 2-dimensional taxonomy of the Realities, along the 2 axes of Physicality (Atoms), and Virtuality (Bits), in which the Realities are shown as children's letter-blocks located with DR at the origin, PR along the Physicality axis, VR along the Virtuality axis, etc..}
    \caption{The Cyber-Physical (Atoms-Bits) Plane has Physical Reality (PR) on the Physicality ("Atoms") axis and Virtual Reality (VR) on the Virtuality ("Bits") axis.  Here the plane is shown segmented into four quadrants.}
    \label{fig:CyberPhysicalPlane}
\end{figure}

For simplicity, the Cyber-Physical Plane alone, in two-dimensions, is shown in Fig.~\ref{fig:CyberPhysicalPlane}

As we trace a path such as a line or arc from VR (Virtuality) to PR (Physicality), we can seamlessly move along a continuum, as suggested by Milgram~\cite{milgram1994taxonomy} and others~\cite{skarbez2021revisiting}.  Milgram's Mixed Reality continuum provides a provides a one-dimensional transition between Physical Reality (PR) and Virtual Reality (VR). However, a significant drawback of this one-dimensional continuum is the lack of origin or zero-point.
A zero-point helps us address important technologies of attenuation, such as dark sunglasses, welding helmets, light shields such as baseball caps, and the like.
More generally, diminished reality is as important as augmented reality~\cite{mann2002eyetap, mann2003introduction, cheng2022towards, mori2017survey, herling2012pixmix, kawai2015diminished, siltanen2017diminished, chen2024diminished, meerits2015real}.
Thus having an axis with an origin is quite a bit more useful than having an axis that has no origin~\cite{mann2023extended}.

Other axis-based taxonomies include the three-dimensional taxonomy (3 axes) of Skarbez~\cite{skarbez2021revisiting} which suffers from a similar problem, lacking a clearly defined zero-point origin for the three axes.

\begin{figure*}
    \centering
    \includegraphics[width=\linewidth]{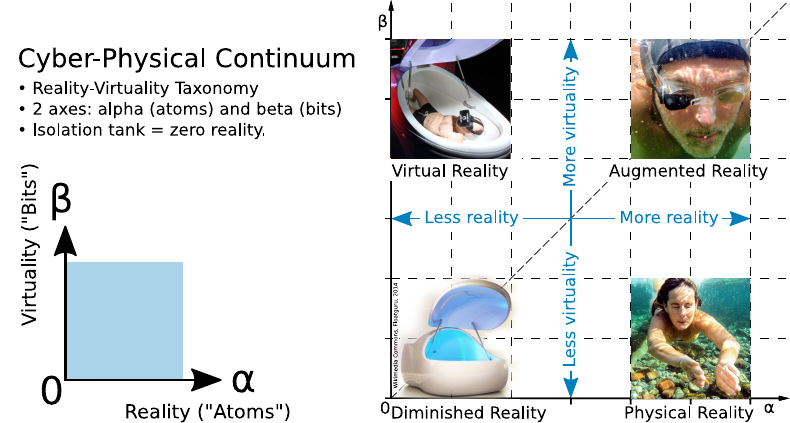}
    \Description[Cyber-Physical Continuum diagram.]{Cyber-Physical Continuum illustrated as axes, left-to-right Reality ("Atoms") and bottom-to-top Virtuality ("Bits").
    This divides the page into 4 quadrants.  At the lower left is a photograph of a float tank.  At the upper left is a photograph of a man in a float tank wearing a VR headset.  At the upper right is a photograph of a man swimming with an XR glass.  At the lower right is a photograph of a woman swimming underwater to represent the physical world without technology.}
    \caption{The Cyber-Physical Continuum with four examples: Icewater swimming gives a strong connection to physical reality (PR).
    A sensory-deprivation float tank diminishes stimulus from physical and virtual sources.  A VR float tank cuts out the physical world so that the only stimulus is the virtual content.  Icewater swimming with a Vuzix SmartSwim provides some physical and some virtual reality content.}
    \label{fig:DR}
\end{figure*}

Taxonomies and ontologies often help identify underexplored areas of research.  Referring to Fig.~\ref{fig:CyberPhysicalPlane} we see the lower-left quadrant, i.e. the zero-point, missing in the Milgram and Skarbez taxonomies, is an underexplored area in the literature on ``the Realities''.  Technologies like dark sunglasses, welding glass, earplugs, etc., are important, and indeed we can innovate with adaptive noise cancellation, smart sunglasses, smart welding helmets, etc., to create a whole new area of research called ``Diminished Reality'' (``DR'')~\cite{mann2002eyetap, mann2003introduction, cheng2022towards, mori2017survey, herling2012pixmix, kawai2015diminished, siltanen2017diminished, chen2024diminished, meerits2015real}, as suggested in Fig.~\ref{fig:DR}.

An example of DR is a sensory-deprivation float-tank where stimulus from reality and virtuality is greatly attenuated.  This doesn't mean the brain does nothing.  In fact we hallucinate strongly in a float tank.  But the input to the senses is close to zero physically (no sound or light, etc.) and virtually (no distracting emails or texts).

The history of Cyber-Physical Systems (CPS) is very nicely summarized by Lee\cite{lee2015past}:
\begin{quote}
  The term “cyber-physical systems” emerged around 2006, when it was coined by Helen Gill at the National Science Foundation in the United States. The related term “cyberspace” is attributed to William Gibson, who used the term in the novel Neuromancer, but the roots of the term CPS are older and deeper. It would be more accurate to view the terms “cyberspace” and “cyber-physical systems” as stemming from the same root, “cybernetics,” which was coined by Norbert Wiener\cite{wiener2019cybernetics}, an American mathematician. Wiener's control logic was effectively a computation, albeit one carried out with analog circuits and mechanical parts, and therefore, cybernetics is the conjunction of physical processes, computation and communication. Wiener derived the term from the Greek κυβερνήτης (kybernetes), meaning helmsman, governor, pilot or rudder.\cite{lee2015past}
\end{quote}

\subsection{HCI and Metaverse (``Bits'' and ``Genes'')}
Human-Computer Interaction (HCI) concerns itself with the interplay between the virtual/cyber world of computing technology (the green ``bits'' axis of Fig.~\ref{fig:teaser}) and the human/social world, which is a third axis denoted in red, as ``Sociality''. These axes (red and green) define the yellow plane (yellow is a color comprised of equal parts red and green) of Fig.~\ref{fig:teaser}.

The concept of ``cyborg'' (a word coined by Manfred Clynes in the 1960s meaning ``cybernetic organism'') also exists in the yellow Socio-Cyber plane denoted ``Metaverse'' (i.e. collaborative social VR).

The IEEE's tagline, ``Advancing Techology for Humanity'', could be construed as also existing in the yellow Metaverse plane.

The cellphone is an example of a technology that can truly connect humans on a social level.
The first handheld wireless telephone, a prototype DynaTAC model (weighting about 2 kg), was invented and used in New York City, 1973 April 3rd, by Martin Cooper of Motorola~\cite{martycoopermobilephone}.


Indeed, social media, often associated with handheld devices like cellphones, also exists in the yellow Cyber-Social plane.

Increasingly, smartphones now support 3D graphics applications as well, giving us shared virtual reality experiences.

\subsection{Integral Ecology (``Atoms'' and ``Genes'')}
Discussion of the Socio-Cyber-Physical space would not be complete without also considering the magenta-colored plane at the bottom of Fig~\ref{fig:teaser}.
Magenta is a colour comprised of red and blue, as this plane is spanned by the blue Physical Reality (``Atoms'') and the red Sociality (``Genes'') axes.

This plane is best described as the plane of ``Integral Ecology''.  Integral Ecology is the study of the relationships between living organisms and their environment~\cite{mickey2013quest}.  The concept was adopted by Pope Francis (born Jorge Mario Bergoglio) in his encyclical Laudato si' (2015)~\cite{si2015care} which emphasizes the interdependence between people and their natural environment, with emphasis on how soil quality, water purification, and air quality affect and are affected by human life.

Integral Ecology seeks to protect both the environment and the well-being of society.
This approach combines environmental sustainability and socio-economic issues.

Francis writes:
``[t]he global economic crises have made painfully obvious the detrimental effects of disregarding our common destiny, which cannot exclude those who come after us.''

Others such as Thomas Berry emphasize harmony between humans adn earth~\cite{mickey2013quest}.

\subsection{Atoms, Bits, and Genes}
The commonly used ``Bits and Atoms'' metaphor (e.g. MIT's Center for Bits and Atoms) is a simplification that takes quite a bit of poetic license (there exist sub-atomic particles, and also computation cannot exist without physical hardware):  ``Bits'' can't exist without ``Atoms''.  Likewise, humans are also made of atoms, but we apply the same poetic license in considering Humanity/Sociality as a separate axis, giving us Atoms, Bits, and Genes.  We call this space the Socio-Cyber-Physical Space, or simply XR-space.

XR-space is the space fully spanned by XR (eXtended Reality), which is the generalized reality that interpolates between the various Realities and eXtrapolates beyond them~\protect\cite{mannwyckoff91, mann2001can, foster2024virtual, simge2024ingilizce, boffi2024co, hoffmann2024verschmelzen}, through technologies that deeply connect to both the physical and social worlds, with such concepts as Immersive XR (e.g. WaterHCI = Water-Human-Computer Interaction)~\cite{mann2021water, pell2023developing, mueller2024grand}.

\section{XR}
Summarizing Fig.~\ref{fig:teaser}.
The XR Space is a taxonomy or ontology of the `R's (``Realities''), the Metaverse, HCI, and the like.
VR (Virtual Reality) exists entirely along the Virtuality axis.  PR (Physical Reality), i.e. the real world, exists along the Reality axis. AR (Augmented Reality) exists in the Reality+Virtuality plane.

Technologies like ear plugs, dark sunglasses, welding helmets, baseball caps, etc., are examples of DR (Diminished Reality) which exists near the origin of the space (where the axes meet). Perhaps the best example of a technology at the origin of XRspace is the sensor deprivation ``float tank'' that provides us with zero stimulus.

HCI (Human-Computer Interaction) and the Metaverse (shared social VR) exist in the Cyber-Social (Sociality and Virtuality) plane.

XR (eXtended Reality) spans the entire Socio-Cyber-Physical space, interpolating between the various Realities and extrapolating beyond them.  What is described here is a simplified version of much of the work outlined in a previous IEEE Special Issue~\protect\cite{el2024metaverse}.

\section{XRscalespace}
The XRspace of Fig.~\ref{fig:teaser} as an organizational taxonomy considers the amount of Physicality, Virtuality, and Sociality, whereas we wish to also consider the scale, and thus introduce XRscalespace in which the three axes are the scale of Physicality, $s(\alpha)$, the scale of Virtuality, $s(\beta)$, and the scale of Sociality, $s(\gamma)$, as in~\cite{mann2014toposculpting} or reciprocal scale as in~\cite{mann2001can}.
If we use logarithmic axes, XRscalespace divides the three-dimensional $s(\alpha), s(\beta), s(\gamma)$ space into 8 octants, as: small v. large physical scale (e.g. clothes versus car versus smart building, smart city, etc.); ``little data'' (distributed e.g. blockchain) v. ``big data'' (centralized),
and sousveillance~\cite{fletcher2011, michael2012sousveillance, bradwell2012security, mann2002sousveillance, bakir2010sousveillance, Freshwater2013Revisiting, Weston2009Embracing, ali2013inevitability, mann2012glassage, minsky2013society, ganascia2010generalized, vertegaal2008attentive, ali2013measuring, nolan2008sousveillance, ali2013comparametric, weber2012surveillance, quessada2010sousveillance, manders2013moving, mannsousveillant, thatcher2017sousveillant, ceccato2019eyes, negativesousveillance,
 measuringVeillanceVixels} (sensing at small social scale as with personal technologies and self-recording) v. surveillance (sensing at large social scale).
It has been observed that most technologies exist along the diagonal of this space, e.g. wearable technolgies (technologies at a small physical scale) tend to be distributed (``little data'')
and sousveillant, whereas city-scale technologies (e.g. technology at a large physical cale) tend to be centralized (``big data'') and surveillant~\cite{mann2001can}.

\section{Future research directions}
The Socio-Cyber-Physical Space/Taxonomy (the ``Atoms-Bits-Genes'' framework) lends itself to a family of taxonomies and ontologies that allow us to organize and understand a wide range of technologies and identify new underexplored areas of possible research, such as city-scale sousveillance, which exist in the ``corners'' and ``edge-cases'' of the taxonomy (e.g. off the main diagonal).
Additionally, much needs to be done to explore areas close to the origin of the space, i.e. Diminished Reality (DR).  For example, how can an AI-eyeglass help us simplify the world and help a visually impaired person find their way, or help someone suffering from excess stimulus (e.g. someone on the autism spectrum) get a simplified understanding of an otherwise overwhelming world.

Moreover, how can we combine sparse areas of the taxonomy space?  For example, how might we extend our senses way out onto the extremes (like more than 100\% of what we can normally sense) while at the same time diminishing other aspects of our senses to make ``space'' for the extra-senses.
How might we use synthetic synesthesia to endow ourselves with additional senses that we can make space for through the power of Diminshed Reality (DR)?

Finally, how can we use all of this to advance technology for humanity and earth (+water + air), e.g. can we tame the monster of technology with a piece of itself, using DR to connect us to our natural world?
Can we create a new reality in which we feel the pain of our actions upon the planet before we take these actions or at least before it is too late?

\section{Conclusion}
This paper highlights the importance of, and key features of, XR = eXtended Reality, as embodied by technologies like the AI-eyeglass.

XR exists within the Socio-Cyber-Physical Space (XR-Space) taxonomy/ontology, which leads us naturally to the concept of Mersivity and six desiderata for technology in service of people and planet.
Technology should be:
\begin{enumerate}
    \item Unmonopolizing so we can continue to sense our surroundings;
    \item Unrestrictive so we can continue to affect our surroundings;
    \item Observable so we can sense and understand the technology;
    \item Controllable so it senses us in a way that serves us and our environment;
    \item Attentive to its environment, e.g. to help us understand our surroundings;
    \item Communicative to our surroundings.  We are part of each others' environment, so in this way the technology serves as a form of communication that truly connects us to each other.
\end{enumerate}

\section{Acknowledgements}
Authors wish to acknowledge many great suggestions by Peter Corcoran, Nishant Kumar, and Mark Billinghurst.

\bibliography{xr}


\begin{thebibliography}{64}


\ifx \showCODEN    \undefined \def \showCODEN     #1{\unskip}     \fi
\ifx \showDOI      \undefined \def \showDOI       #1{#1}\fi
\ifx \showISBNx    \undefined \def \showISBNx     #1{\unskip}     \fi
\ifx \showISBNxiii \undefined \def \showISBNxiii  #1{\unskip}     \fi
\ifx \showISSN     \undefined \def \showISSN      #1{\unskip}     \fi
\ifx \showLCCN     \undefined \def \showLCCN      #1{\unskip}     \fi
\ifx \shownote     \undefined \def \shownote      #1{#1}          \fi
\ifx \showarticletitle \undefined \def \showarticletitle #1{#1}   \fi
\ifx \showURL      \undefined \def \showURL       {\relax}        \fi
\providecommand\bibfield[2]{#2}
\providecommand\bibinfo[2]{#2}
\providecommand\natexlab[1]{#1}
\providecommand\showeprint[2][]{arXiv:#2}

\bibitem[cam({[n.\,d.]})]%
        {campuscanada}
 \bibinfo{year}{[n.\,d.]}\natexlab{}.
\newblock \showarticletitle{Steve Mann}.
\newblock \bibinfo{journal}{\emph{Campus Canada, ISSN 0823-4531, p55 Feb-Mar 1985, pp58-59 Apr-May 1986, p72 Sep-Oct 1986}} (\bibinfo{year}{[n.\,d.]}).
\newblock


\bibitem[imp(1985)]%
        {impulse}
 \bibinfo{year}{October 1985}\natexlab{}.
\newblock \showarticletitle{Impulse}.
\newblock  \bibinfo{volume}{12}, \bibinfo{number}{2} (\bibinfo{year}{October 1985}).
\newblock


\bibitem[Ali et~al\mbox{.}(2013a)]%
        {ali2013comparametric}
\bibfield{author}{\bibinfo{person}{Mir~Adnan Ali}, \bibinfo{person}{Tao Ai}, \bibinfo{person}{Akshay Gill}, \bibinfo{person}{Jose Emilio}, \bibinfo{person}{Kalin Ovtcharov}, {and} \bibinfo{person}{Steve Mann}.} \bibinfo{year}{2013}\natexlab{a}.
\newblock \showarticletitle{Comparametric {HDR} ({H}igh {D}ynamic {R}ange) imaging for digital eye glass, wearable cameras, and sousveillance}. In \bibinfo{booktitle}{\emph{ISTAS}}. IEEE, \bibinfo{pages}{107--114}.
\newblock


\bibitem[Ali and Mann(2013)]%
        {ali2013inevitability}
\bibfield{author}{\bibinfo{person}{Mir~Adnan Ali} {and} \bibinfo{person}{Steve Mann}.} \bibinfo{year}{2013}\natexlab{}.
\newblock \showarticletitle{The inevitability of the transition from a surveillance-society to a veillance-society: Moral and economic grounding for sousveillance}. In \bibinfo{booktitle}{\emph{ISTAS}}. IEEE, \bibinfo{pages}{243--254}.
\newblock


\bibitem[Ali et~al\mbox{.}(2013b)]%
        {ali2013measuring}
\bibfield{author}{\bibinfo{person}{Mir~Adnan Ali}, \bibinfo{person}{Jonathan~Polak Nachumow}, \bibinfo{person}{Jocelyn~A Srigley}, \bibinfo{person}{Colin~D Furness}, \bibinfo{person}{Sebastian Mann}, {and} \bibinfo{person}{Michael Gardam}.} \bibinfo{year}{2013}\natexlab{b}.
\newblock \showarticletitle{Measuring the effect of sousveillance in increasing socially desirable behaviour}. In \bibinfo{booktitle}{\emph{ISTAS}}. IEEE, \bibinfo{pages}{266--267}.
\newblock


\bibitem[Artaud(1958)]%
        {artaud}
\bibfield{author}{\bibinfo{person}{Antonin Artaud}.} \bibinfo{year}{1958}\natexlab{}.
\newblock \bibinfo{title}{The Theater and Its Double}.
\newblock , \bibinfo{numpages}{159}~pages.
\newblock
\newblock
\shownote{first published reference to ``la realite virtuell'' (``virtual reality'')}.


\bibitem[Bakir(2010)]%
        {bakir2010sousveillance}
\bibfield{author}{\bibinfo{person}{V. Bakir}.} \bibinfo{year}{2010}\natexlab{}.
\newblock \bibinfo{booktitle}{\emph{Sousveillance, media and strategic political communication}}.
\newblock \bibinfo{publisher}{Continuum}.
\newblock


\bibitem[Boffi and Boccia(2024)]%
        {boffi2024co}
\bibfield{author}{\bibinfo{person}{Laura Boffi} {and} \bibinfo{person}{Gianluca Boccia}.} \bibinfo{year}{2024}\natexlab{}.
\newblock \showarticletitle{The Co-Drive Service: Probing into Preliminary Social Outcomes through Early-Stage Prototyping}.
\newblock  (\bibinfo{year}{2024}).
\newblock


\bibitem[Bradwell and Michael(2012)]%
        {bradwell2012security}
\bibfield{author}{\bibinfo{person}{J. Bradwell} {and} \bibinfo{person}{K. Michael}.} \bibinfo{year}{2012}\natexlab{}.
\newblock \showarticletitle{Security workshop brings `sousveillance' under the microscope}.
\newblock \bibinfo{journal}{\emph{University of Wollongong: Latest News}} (\bibinfo{year}{2012}).
\newblock
\newblock
\shownote{\url{http://media.uow.edu.au/news/UOW120478.html}, accessed 2014}.


\bibitem[Ceccato(2019)]%
        {ceccato2019eyes}
\bibfield{author}{\bibinfo{person}{Vania Ceccato}.} \bibinfo{year}{2019}\natexlab{}.
\newblock \showarticletitle{Eyes and Apps on the Streets: From Surveillance to Sousveillance Using Smartphones}.
\newblock \bibinfo{journal}{\emph{Criminal Justice Review}} \bibinfo{volume}{44}, \bibinfo{number}{1} (\bibinfo{year}{2019}), \bibinfo{pages}{25--41}.
\newblock


\bibitem[Chen et~al\mbox{.}(2024)]%
        {chen2024diminished}
\bibfield{author}{\bibinfo{person}{Siru Chen}, \bibinfo{person}{Lingxin Yu}, \bibinfo{person}{Yuxuan Liu}, \bibinfo{person}{Zhifei Ding}, \bibinfo{person}{Jiacheng Zhang}, \bibinfo{person}{Xinyue Wang}, \bibinfo{person}{Jiahao Han}, {and} \bibinfo{person}{Richen Liu}.} \bibinfo{year}{2024}\natexlab{}.
\newblock \showarticletitle{Diminished reality techniques for metaverse applications: A perspective from evaluation}.
\newblock \bibinfo{journal}{\emph{IEEE Internet of Things Journal}} (\bibinfo{year}{2024}).
\newblock


\bibitem[Cheng et~al\mbox{.}(2022)]%
        {cheng2022towards}
\bibfield{author}{\bibinfo{person}{Yi~Fei Cheng}, \bibinfo{person}{Hang Yin}, \bibinfo{person}{Yukang Yan}, \bibinfo{person}{Jan Gugenheimer}, {and} \bibinfo{person}{David Lindlbauer}.} \bibinfo{year}{2022}\natexlab{}.
\newblock \showarticletitle{Towards understanding diminished reality}. In \bibinfo{booktitle}{\emph{Proceedings of the 2022 CHI Conference on Human Factors in Computing Systems}}. \bibinfo{pages}{1--16}.
\newblock


\bibitem[El~Saddik et~al\mbox{.}(2024)]%
        {el2024metaverse}
\bibfield{author}{\bibinfo{person}{Abdulmotaleb El~Saddik}, \bibinfo{person}{Fabrizio Lamberti}, \bibinfo{person}{Steve Mann}, \bibinfo{person}{Filippo~Gabriele Prattic{\`o}}, \bibinfo{person}{Ruck Thawonmas}, {and} \bibinfo{person}{Yu Yuan}.} \bibinfo{year}{2024}\natexlab{}.
\newblock \showarticletitle{Metaverse and eXtended uniVerse (XV): Opportunities and Challenges for Consumer Technologies}.
\newblock \bibinfo{journal}{\emph{IEEE Consumer Electronics Magazine}} (\bibinfo{year}{2024}).
\newblock


\bibitem[Fletcher et~al\mbox{.}(2011)]%
        {fletcher2011}
\bibfield{author}{\bibinfo{person}{G. Fletcher}, \bibinfo{person}{M. Griffiths}, {and} \bibinfo{person}{M. Kutar}.} \bibinfo{year}{September 7, 2011}\natexlab{}.
\newblock \showarticletitle{A day in the digital life: a preliminary sousveillance study}.
\newblock \bibinfo{journal}{\emph{SSRN, http://papers.ssrn.com/sol3/papers.cfm?abstract\_id=1923629}} (\bibinfo{year}{September 7, 2011}).
\newblock


\bibitem[Foster et~al\mbox{.}(2024)]%
        {foster2024virtual}
\bibfield{author}{\bibinfo{person}{Sophie Foster}, \bibinfo{person}{Larissa Barth}, {and} \bibinfo{person}{Zaryab Chaudhry}.} \bibinfo{year}{2024}\natexlab{}.
\newblock \showarticletitle{Virtual Gathering Platforms in Academic Teaching: Potential and Applications}.
\newblock \bibinfo{journal}{\emph{Electronic Journal of e-Learning}} \bibinfo{volume}{22}, \bibinfo{number}{3} (\bibinfo{year}{2024}), \bibinfo{pages}{124--140}.
\newblock


\bibitem[Freshwater et~al\mbox{.}(2013)]%
        {Freshwater2013Revisiting}
\bibfield{author}{\bibinfo{person}{D. Freshwater}, \bibinfo{person}{P. Fisher}, {and} \bibinfo{person}{E. Walsh}.} \bibinfo{year}{2013}\natexlab{}.
\newblock \showarticletitle{Revisiting the Panopticon: professional regulation, surveillance and sousveillance}.
\newblock \bibinfo{journal}{\emph{Nursing Inquiry}} (\bibinfo{date}{May} \bibinfo{year}{2013}).
\newblock
\showISSN{1440-1800}
\urldef\tempurl%
\url{https://doi.org/10.1111/nin.12038}
\showDOI{\tempurl}
\newblock
\shownote{PMID: 23718546}.


\bibitem[Ganascia(2010)]%
        {ganascia2010generalized}
\bibfield{author}{\bibinfo{person}{Jean-Gabriel Ganascia}.} \bibinfo{year}{2010}\natexlab{}.
\newblock \showarticletitle{The generalized sousveillance society}.
\newblock \bibinfo{journal}{\emph{Social Science Information}} \bibinfo{volume}{49}, \bibinfo{number}{3} (\bibinfo{year}{2010}), \bibinfo{pages}{489--507}.
\newblock


\bibitem[Herling and Broll(2012)]%
        {herling2012pixmix}
\bibfield{author}{\bibinfo{person}{Jan Herling} {and} \bibinfo{person}{Wolfgang Broll}.} \bibinfo{year}{2012}\natexlab{}.
\newblock \showarticletitle{Pixmix: A real-time approach to high-quality diminished reality}. In \bibinfo{booktitle}{\emph{2012 ieee international symposium on mixed and augmented reality (ismar)}}. IEEE, \bibinfo{pages}{141--150}.
\newblock


\bibitem[Hoffmann(2024)]%
        {hoffmann2024verschmelzen}
\bibfield{author}{\bibinfo{person}{Peter Hoffmann}.} \bibinfo{year}{2024}\natexlab{}.
\newblock \showarticletitle{Das Verschmelzen von Welten und… versen}.
\newblock In \bibinfo{booktitle}{\emph{Next Generation Internet: Die Verschmelzung von Realit{\"a}t und Virtualit{\"a}t im Metaversum}}. \bibinfo{publisher}{Springer}, \bibinfo{pages}{27--86}.
\newblock


\bibitem[Janzen and Mann(2014)]%
        {measuringVeillanceVixels}
\bibfield{author}{\bibinfo{person}{Ryan. Janzen} {and} \bibinfo{person}{Steve. Mann}.} \bibinfo{year}{2014}\natexlab{}.
\newblock \showarticletitle{Vixels, Veillons, Veillance Flux: An extramissive information-bearing formulation of sensing, to measure Surveillance and Sousveillance}.
\newblock \bibinfo{journal}{\emph{IEEE CCECE}} (\bibinfo{year}{2014}), \bibinfo{pages}{1--10}.
\newblock


\bibitem[Kawai et~al\mbox{.}(2015)]%
        {kawai2015diminished}
\bibfield{author}{\bibinfo{person}{Norihiko Kawai}, \bibinfo{person}{Tomokazu Sato}, {and} \bibinfo{person}{Naokazu Yokoya}.} \bibinfo{year}{2015}\natexlab{}.
\newblock \showarticletitle{Diminished reality based on image inpainting considering background geometry}.
\newblock \bibinfo{journal}{\emph{IEEE transactions on visualization and computer graphics}} \bibinfo{volume}{22}, \bibinfo{number}{3} (\bibinfo{year}{2015}), \bibinfo{pages}{1236--1247}.
\newblock


\bibitem[Lee(2015)]%
        {lee2015past}
\bibfield{author}{\bibinfo{person}{Edward~A Lee}.} \bibinfo{year}{2015}\natexlab{}.
\newblock \showarticletitle{The past, present and future of cyber-physical systems: A focus on models}.
\newblock \bibinfo{journal}{\emph{Sensors}} \bibinfo{volume}{15}, \bibinfo{number}{3} (\bibinfo{year}{2015}), \bibinfo{pages}{4837--4869}.
\newblock


\bibitem[Manders(2013)]%
        {manders2013moving}
\bibfield{author}{\bibinfo{person}{Corey Manders}.} \bibinfo{year}{2013}\natexlab{}.
\newblock \showarticletitle{Moving surveillance techniques to sousveillance: Towards equiveillance using wearable computing}. In \bibinfo{booktitle}{\emph{ISTAS}}. IEEE, \bibinfo{pages}{19--19}.
\newblock


\bibitem[{Mann}(1992)]%
        {mannws}
\bibfield{author}{\bibinfo{person}{S. {Mann}}.} \bibinfo{year}{1992}\natexlab{}.
\newblock \showarticletitle{Wavelets and "Chirplets": Time--Frequency "Perspectives" With Applications}.
\newblock In \bibinfo{booktitle}{\emph{Advances in machine vision: strategies and applications}}. \bibinfo{publisher}{World Scientific}, \bibinfo{pages}{99--128}.
\newblock


\bibitem[Mann(1997a)]%
        {mannars}
\bibfield{author}{\bibinfo{person}{Steve Mann}.} \bibinfo{year}{1997}\natexlab{a}.
\newblock \showarticletitle{{H}umanistic {I}ntelligence ({H.I.})}.
\newblock \bibinfo{journal}{\emph{Proceedings of Ars Electronica}} (\bibinfo{date}{Sep 8-13} \bibinfo{year}{1997}), \bibinfo{pages}{217--231}.
\newblock
\newblock
\shownote{Invited plenary lecture, Sep. 10, http://wearcam.org/ars/ http//www.aec.at/fleshfactor, Republished in: Timothy Druckrey (ed.), Ars Electronica: Facing the Future, A Survey of Two Decades, MIT Press, pp 420--427}.


\bibitem[Mann(1997b)]%
        {mannarsperformance}
\bibfield{author}{\bibinfo{person}{Steve Mann}.} \bibinfo{year}{1997}\natexlab{b}.
\newblock \showarticletitle{Sicherheitsglaeser}.
\newblock \bibinfo{journal}{\emph{Public performance at Ars Electronica}} (\bibinfo{date}{Sep 8-13} \bibinfo{year}{1997}).
\newblock
\newblock
\shownote{See also http://wearcam.org/ars/ and http://wearcam.org/lvac/ where it was also exhibited at List Visual Arts Centre, Oct9 - Dec28, 1997}.


\bibitem[Mann(2001)]%
        {mann2001can}
\bibfield{author}{\bibinfo{person}{Steve Mann}.} \bibinfo{year}{2001}\natexlab{}.
\newblock \showarticletitle{Can humans being clerks make clerks be human?--exploring the fundamental difference between ubicomp and wearcomp (k{\"o}nnen menschen, die sich wie angestellte benehmen, angestellte zu menschlichem verhalten bewegen? zum fundamentalen unterschied zwischen ubicomp und wearcomp)}.
\newblock \bibinfo{journal}{\emph{it-Information Technology}} \bibinfo{volume}{43}, \bibinfo{number}{2} (\bibinfo{year}{2001}), \bibinfo{pages}{97--106}.
\newblock


\bibitem[Mann(2002)]%
        {mann2002sousveillance}
\bibfield{author}{\bibinfo{person}{Steve Mann}.} \bibinfo{year}{2002}\natexlab{}.
\newblock \showarticletitle{Sousveillance, not just surveillance, in response to terrorism}.
\newblock \bibinfo{journal}{\emph{Metal and Flesh}} \bibinfo{volume}{6}, \bibinfo{number}{1} (\bibinfo{year}{2002}), \bibinfo{pages}{1--8}.
\newblock


\bibitem[{Mann}(2016)]%
        {kineveillance}
\bibfield{author}{\bibinfo{person}{S. {Mann}}.} \bibinfo{year}{2016}\natexlab{}.
\newblock \showarticletitle{Surveillance (Oversight), Sousveillance (Undersight), and Metaveillance (Seeing Sight Itself)}. In \bibinfo{booktitle}{\emph{2016 IEEE Conference on Computer Vision and Pattern Recognition Workshops (CVPRW)}}. \bibinfo{pages}{1408--1417}.
\newblock


\bibitem[Mann(2018)]%
        {mann2018phenomenological}
\bibfield{author}{\bibinfo{person}{S. Mann}.} \bibinfo{year}{IEEE GEM2018}\natexlab{}.
\newblock \showarticletitle{Phenomenological Augmented Reality with the Sequential Wave Imprinting Machine (SWIM)}. In \bibinfo{booktitle}{\emph{2018 IEEE Games, Entertainment, Media Conference (GEM)}}. \bibinfo{pages}{220--227}.
\newblock


\bibitem[Mann(1998)]%
        {soundfieldshirt}
\bibfield{author}{\bibinfo{person}{Steve Mann}.} \bibinfo{year}{Night Gallery, 185 Richmond Street West, Toronto, Ontario, Canada, July 1985. Later exhibited at Hamilton Artists Inc, 1998}\natexlab{}.
\newblock \showarticletitle{Wearable Technologies}.
\newblock  (\bibinfo{year}{Night Gallery, 185 Richmond Street West, Toronto, Ontario, Canada, July 1985. Later exhibited at Hamilton Artists Inc, 1998}).
\newblock


\bibitem[Mann(2012)]%
        {mann2012glassage}
\bibfield{author}{\bibinfo{person}{S. Mann}.} \bibinfo{year}{{Nov. 02, 2012}}\natexlab{}.
\newblock \showarticletitle{Eye Am a Camera: Surveillance and Sousveillance in the Glassage}.
\newblock \bibinfo{journal}{\emph{TIME}} (\bibinfo{year}{{Nov. 02, 2012}}).
\newblock


\bibitem[Mann and Barfield(2003)]%
        {mann2003introduction}
\bibfield{author}{\bibinfo{person}{Steve Mann} {and} \bibinfo{person}{Woodrow Barfield}.} \bibinfo{year}{2003}\natexlab{}.
\newblock \showarticletitle{Introduction to mediated reality}.
\newblock \bibinfo{journal}{\emph{International Journal of Human-Computer Interaction}} \bibinfo{volume}{15}, \bibinfo{number}{2} (\bibinfo{year}{2003}), \bibinfo{pages}{205--208}.
\newblock


\bibitem[Mann et~al\mbox{.}(2001)]%
        {wearableai}
\bibfield{author}{\bibinfo{person}{Steve Mann}, \bibinfo{person}{Li-Te Cheng}, \bibinfo{person}{John Robinson}, \bibinfo{person}{Kaoru Sumi}, \bibinfo{person}{Toyoaki Nishida}, \bibinfo{person}{Soichiro Matsushita}, \bibinfo{person}{Ömer Faruk~Özer}, \bibinfo{person}{Oˇguz Özün}, \bibinfo{person}{C. Öncel Tüzel}, \bibinfo{person}{Volkan Atalay}, \bibinfo{person}{A.~Enis Çetin}, \bibinfo{person}{Joshua Anhalt}, \bibinfo{person}{Asim Smailagic}, \bibinfo{person}{Daniel~P. Siewiorek}, \bibinfo{person}{Francine Gemperle}, \bibinfo{person}{Daniel Salber}, \bibinfo{person}{Sam Weber}, \bibinfo{person}{Jim Beck}, \bibinfo{person}{Jim Jennings}, {and} \bibinfo{person}{David~A. Ross}.} \bibinfo{year}{May/June 2001}\natexlab{}.
\newblock \showarticletitle{Wearable AI}.
\newblock \bibinfo{journal}{\emph{Intelligent Systems, IEEE}} \bibinfo{volume}{16}, \bibinfo{number}{3} (\bibinfo{year}{May/June 2001}), \bibinfo{pages}{1--53}.
\newblock


\bibitem[Mann et~al\mbox{.}(2020)]%
        {mann2020ichms}
\bibfield{author}{\bibinfo{person}{Steve Mann}, \bibinfo{person}{Phillip~V. Do}, \bibinfo{person}{Danson~Evan Garcia}, \bibinfo{person}{Jesse Hernandez}, {and} \bibinfo{person}{Humza Khokhar}.} \bibinfo{year}{2020}\natexlab{}.
\newblock \showarticletitle{Electrical Engineering Design with the Subconscious Mind}. In \bibinfo{booktitle}{\emph{2020 1st IEEE International Conference on Human-Machine Systems (ICHMS)}}. \bibinfo{pages}{1--6}.
\newblock


\bibitem[Mann and Fung(2002)]%
        {mann2002eyetap}
\bibfield{author}{\bibinfo{person}{Steve Mann} {and} \bibinfo{person}{James Fung}.} \bibinfo{year}{2002}\natexlab{}.
\newblock \showarticletitle{EyeTap devices for augmented, deliberately diminished, or otherwise altered visual perception of rigid planar patches of real-world scenes}.
\newblock \bibinfo{journal}{\emph{Presence}} \bibinfo{volume}{11}, \bibinfo{number}{2} (\bibinfo{year}{2002}), \bibinfo{pages}{158--175}.
\newblock


\bibitem[Mann et~al\mbox{.}(2018)]%
        {mannsousveillant}
\bibfield{author}{\bibinfo{person}{S. Mann}, \bibinfo{person}{J.~C. Havens}, \bibinfo{person}{G. Cowan}, \bibinfo{person}{A. Richardson}, {and} \bibinfo{person}{R. Ouellette}.} \bibinfo{year}{2018}\natexlab{}.
\newblock \showarticletitle{Sousveillant cities and media}.
\newblock \bibinfo{journal}{\emph{Mesh Cities}} (\bibinfo{year}{2018}), \bibinfo{pages}{1--15}.
\newblock


\bibitem[Mann et~al\mbox{.}(2014)]%
        {mann2014toposculpting}
\bibfield{author}{\bibinfo{person}{Steve Mann}, \bibinfo{person}{Ryan Janzen}, \bibinfo{person}{Tao Ai}, \bibinfo{person}{Seyed~Nima Yasrebi}, \bibinfo{person}{Jad Kawwa}, {and} \bibinfo{person}{Mir~Adnan Ali}.} \bibinfo{year}{2014}\natexlab{}.
\newblock \showarticletitle{Toposculpting: Computational lightpainting and wearable computational photography for abakographic user interfaces}. In \bibinfo{booktitle}{\emph{2014 IEEE 27th Canadian Conference on Electrical and Computer Engineering (CCECE)}}. IEEE, \bibinfo{pages}{1--10}.
\newblock


\bibitem[Mann et~al\mbox{.}(2021)]%
        {mann2021water}
\bibfield{author}{\bibinfo{person}{Steve Mann}, \bibinfo{person}{Mark Mattson}, \bibinfo{person}{Steve Hulford}, \bibinfo{person}{Mark Fox}, \bibinfo{person}{Kevin Mako}, \bibinfo{person}{Ryan Janzen}, \bibinfo{person}{Maya Burhanpurkar}, \bibinfo{person}{Simone Browne}, \bibinfo{person}{Craig Travers}, \bibinfo{person}{Robert Thurmond}, {et~al\mbox{.}}} \bibinfo{year}{2021}\natexlab{}.
\newblock \showarticletitle{Water-Human-Computer-Interface (WaterHCI): Crossing the borders of computation clothes skin and surface}.
\newblock \bibinfo{journal}{\emph{Proceedings of the 23rd annual Water-Human-Computer Interface Deconference (Ontario Place TeachBeach, Toronto, Ontario, Canada). Ontario Place TeachBeach, Toronto, Ontario, Canada}} (\bibinfo{year}{2021}), \bibinfo{pages}{6--35}.
\newblock


\bibitem[Mann and Wyckoff(1991)]%
        {mannwyckoff91}
\bibfield{author}{\bibinfo{person}{Steve Mann} {and} \bibinfo{person}{Charles Wyckoff}.} \bibinfo{year}{1991}\natexlab{}.
\newblock \showarticletitle{Extended Reality}.
\newblock \bibinfo{journal}{\emph{{MIT 4-405}, http://wearcam.org/xr.htm}} (\bibinfo{year}{1991}).
\newblock


\bibitem[Mann et~al\mbox{.}(2023)]%
        {mann2023extended}
\bibfield{author}{\bibinfo{person}{Steve Mann}, \bibinfo{person}{Yu Yuan}, \bibinfo{person}{Fabrizio Lamberti}, \bibinfo{person}{Abdulmotaleb El~Saddik}, \bibinfo{person}{Ruck Thawonmas}, {and} \bibinfo{person}{Filippo~Gabriele Prattico}.} \bibinfo{year}{2023}\natexlab{}.
\newblock \showarticletitle{eXtended meta-uni-omni-Verse (XV): Introduction, Taxonomy, and State-of-the-Art}.
\newblock \bibinfo{journal}{\emph{IEEE Consumer Electronics Magazine}} \bibinfo{volume}{13}, \bibinfo{number}{3} (\bibinfo{year}{2023}), \bibinfo{pages}{27--35}.
\newblock


\bibitem[Meerits and Saito(2015)]%
        {meerits2015real}
\bibfield{author}{\bibinfo{person}{Siim Meerits} {and} \bibinfo{person}{Hideo Saito}.} \bibinfo{year}{2015}\natexlab{}.
\newblock \showarticletitle{Real-time diminished reality for dynamic scenes}. In \bibinfo{booktitle}{\emph{2015 IEEE International Symposium on Mixed and Augmented Reality Workshops}}. IEEE, \bibinfo{pages}{53--59}.
\newblock


\bibitem[Michael and Michael(2012)]%
        {michael2012sousveillance}
\bibfield{author}{\bibinfo{person}{K. Michael} {and} \bibinfo{person}{MG Michael}.} \bibinfo{year}{2012}\natexlab{}.
\newblock \showarticletitle{Sousveillance and Point of View Technologies in Law Enforcement: An Overview}.
\newblock  (\bibinfo{year}{2012}).
\newblock


\bibitem[Mickey et~al\mbox{.}(2013)]%
        {mickey2013quest}
\bibfield{author}{\bibinfo{person}{Sam Mickey}, \bibinfo{person}{Adam Robbert}, {and} \bibinfo{person}{Laura Reddick}.} \bibinfo{year}{2013}\natexlab{}.
\newblock \showarticletitle{The Quest for Integral Ecology.}
\newblock \bibinfo{journal}{\emph{Integral Review: A Transdisciplinary \& Transcultural Journal for New Thought, Research, \& Praxis}} \bibinfo{volume}{9}, \bibinfo{number}{3} (\bibinfo{year}{2013}).
\newblock


\bibitem[Milgram and Kishino(1994)]%
        {milgram1994taxonomy}
\bibfield{author}{\bibinfo{person}{Paul Milgram} {and} \bibinfo{person}{Fumio Kishino}.} \bibinfo{year}{1994}\natexlab{}.
\newblock \showarticletitle{A taxonomy of mixed reality visual displays}.
\newblock \bibinfo{journal}{\emph{IEICE TRANSACTIONS on Information and Systems}} \bibinfo{volume}{77}, \bibinfo{number}{12} (\bibinfo{year}{1994}), \bibinfo{pages}{1321--1329}.
\newblock


\bibitem[Minsky et~al\mbox{.}({[n.\,d.]})]%
        {minsky2013society}
\bibfield{author}{\bibinfo{person}{M. Minsky}, \bibinfo{person}{R. Kurzweil}, {and} \bibinfo{person}{S. Mann}.} \bibinfo{year}{[n.\,d.]}\natexlab{}.
\newblock \showarticletitle{Society of intelligent veillance}. In \bibinfo{booktitle}{\emph{IEEE ISTAS 2013}}.
\newblock


\bibitem[Mori et~al\mbox{.}(2017)]%
        {mori2017survey}
\bibfield{author}{\bibinfo{person}{Shohei Mori}, \bibinfo{person}{Sei Ikeda}, {and} \bibinfo{person}{Hideo Saito}.} \bibinfo{year}{2017}\natexlab{}.
\newblock \showarticletitle{A survey of diminished reality: Techniques for visually concealing, eliminating, and seeing through real objects}.
\newblock \bibinfo{journal}{\emph{IPSJ Transactions on Computer Vision and Applications}}  \bibinfo{volume}{9} (\bibinfo{year}{2017}), \bibinfo{pages}{1--14}.
\newblock


\bibitem[Morrow et~al\mbox{.}(2020)]%
        {cxi2019ieee}
\bibfield{author}{\bibinfo{person}{Monique Morrow}, \bibinfo{person}{Jay Iorio}, \bibinfo{person}{Greg Adamson}, \bibinfo{person}{BC Biermann}, \bibinfo{person}{Katryna Dow}, \bibinfo{person}{Takashi Egawa}, \bibinfo{person}{Danit Gal}, \bibinfo{person}{Ann Greenberg}, \bibinfo{person}{John~C. Havens}, \bibinfo{person}{Dr. Sara~R. Jordan}, \bibinfo{person}{Lauren Joseph}, \bibinfo{person}{Ceyhun Karasu}, \bibinfo{person}{Hyo eun Kim}, \bibinfo{person}{Scott Kesselman}, \bibinfo{person}{Steve Mann}, \bibinfo{person}{Preeti Mohan}, \bibinfo{person}{Lisa Morgan}, \bibinfo{person}{Pablo Noriega}, \bibinfo{person}{Dr.~Stephen Rainey}, \bibinfo{person}{Todd Richard}, \bibinfo{person}{Skip Rizzo}, \bibinfo{person}{Francesca Rossi}, \bibinfo{person}{Leanne Seeto}, \bibinfo{person}{Alan Smithson}, \bibinfo{person}{Mathana Stender}, {and} \bibinfo{person}{Maya Zuckerman}.} \bibinfo{year}{2019-2020}\natexlab{}.
\newblock \showarticletitle{The IEEE global initiative on ethics of autonomous and intelligent systems}.
\newblock \bibinfo{journal}{\emph{Extended Reality in {A/IS}}} (\bibinfo{year}{2019-2020}), \bibinfo{pages}{1--29}.
\newblock


\bibitem[Mueller et~al\mbox{.}(2024)]%
        {mueller2024grand}
\bibfield{author}{\bibinfo{person}{Florian~‘Floyd’ Mueller}, \bibinfo{person}{Maria~F Montoya}, \bibinfo{person}{Sarah~Jane Pell}, \bibinfo{person}{Leif Oppermann}, \bibinfo{person}{Mark Blythe}, \bibinfo{person}{Paul~H Dietz}, \bibinfo{person}{Joe Marshall}, \bibinfo{person}{Scott Bateman}, \bibinfo{person}{Ian Smith}, \bibinfo{person}{Swamy Ananthanarayan}, \bibinfo{person}{Ali Mazalek}, \bibinfo{person}{Alexander Verni}, \bibinfo{person}{Alexander Bakogeorge}, \bibinfo{person}{Mathieu Simonnet}, \bibinfo{person}{Kirsten Ellis}, \bibinfo{person}{Nathan~Arthur Semertzidis}, \bibinfo{person}{Winslow Burleson}, \bibinfo{person}{John Quarles}, \bibinfo{person}{Steve Mann}, \bibinfo{person}{Chris Hill}, \bibinfo{person}{Christal Clashing}, {and} \bibinfo{person}{Don~Samitha Elvitigala}.} \bibinfo{year}{2024}\natexlab{}.
\newblock \showarticletitle{Grand challenges in WaterHCI}. In \bibinfo{booktitle}{\emph{Proceedings of the CHI Conference on Human Factors in Computing Systems}}. \bibinfo{pages}{1--18}.
\newblock


\bibitem[Nolan et~al\mbox{.}(2008)]%
        {nolan2008sousveillance}
\bibfield{author}{\bibinfo{person}{Jason Nolan}, \bibinfo{person}{Steve Mann}, {and} \bibinfo{person}{Barry Wellman}.} \bibinfo{year}{2008}\natexlab{}.
\newblock \showarticletitle{Sousveillance: Wearable and Digital Tools in Surveilled Environments}.
\newblock \bibinfo{journal}{\emph{Small Tech: The Culture of Digital Tools}}  \bibinfo{volume}{22} (\bibinfo{year}{2008}), \bibinfo{pages}{179}.
\newblock


\bibitem[Pell et~al\mbox{.}(2023)]%
        {pell2023developing}
\bibfield{author}{\bibinfo{person}{Sarah~Jane Pell}, \bibinfo{person}{Steve Mann}, {and} \bibinfo{person}{Michael Lombardi}.} \bibinfo{year}{2023}\natexlab{}.
\newblock \showarticletitle{Developing WaterHCI and OceanicXV technologies for diving}. In \bibinfo{booktitle}{\emph{OCEANS 2023-Limerick}}. IEEE, \bibinfo{pages}{1--10}.
\newblock


\bibitem[Quessada(2010)]%
        {quessada2010sousveillance}
\bibfield{author}{\bibinfo{person}{Dominique Quessada}.} \bibinfo{year}{2010}\natexlab{}.
\newblock \showarticletitle{De la sousveillance}.
\newblock \bibinfo{journal}{\emph{Multitudes}} \bibinfo{number}{1} (\bibinfo{year}{2010}), \bibinfo{pages}{54--59}.
\newblock


\bibitem[Reynolds(mark)]%
        {negativesousveillance}
\bibfield{author}{\bibinfo{person}{C. Reynolds}.} \bibinfo{year}{July 4 - 6, 2011, Aarhus, Denmark}\natexlab{}.
\newblock \showarticletitle{Negative Sousveillance}.
\newblock \bibinfo{journal}{\emph{First International Conference of the International Association for Computing and Philosophy (IACAP11)}} (\bibinfo{year}{July 4 - 6, 2011, Aarhus, Denmark}), \bibinfo{pages}{306 -- 309}.
\newblock


\bibitem[Si(2015)]%
        {si2015care}
\bibfield{author}{\bibinfo{person}{Laudato Si}.} \bibinfo{year}{2015}\natexlab{}.
\newblock \showarticletitle{On Care for Our Common Home}.
\newblock \bibinfo{journal}{\emph{London: Catholic Truth Society}} (\bibinfo{year}{2015}).
\newblock


\bibitem[Siltanen(2017)]%
        {siltanen2017diminished}
\bibfield{author}{\bibinfo{person}{Sanni Siltanen}.} \bibinfo{year}{2017}\natexlab{}.
\newblock \showarticletitle{Diminished reality for augmented reality interior design}.
\newblock \bibinfo{journal}{\emph{The Visual Computer}}  \bibinfo{volume}{33} (\bibinfo{year}{2017}), \bibinfo{pages}{193--208}.
\newblock


\bibitem[Simge and Mert(2024)]%
        {simge2024ingilizce}
\bibfield{author}{\bibinfo{person}{U{\u{g}}urluer Simge} {and} \bibinfo{person}{Seven Mert}.} \bibinfo{year}{2024}\natexlab{}.
\newblock \showarticletitle{{\.I}ngilizce ileti{\c{s}}im {\c{c}}al{\i}{\c{s}}malar{\i}nda geni{\c{s}}letilmi{\c{s}} ger{\c{c}}eklik ara{\c{s}}t{\i}rma e{\u{g}}ilimlerinin bibliyometrik bir analizi: Geni{\c{s}}letilmi{\c{s}} ger{\c{c}}eklik teknolojilerine geli{\c{s}}en ra{\u{g}}betin haritalanmas{\i}}.
\newblock \bibinfo{journal}{\emph{Connectist: Istanbul University Journal of Communication Sciences}} \bibinfo{number}{66} (\bibinfo{year}{2024}), \bibinfo{pages}{147--181}.
\newblock


\bibitem[Skarbez et~al\mbox{.}(2021)]%
        {skarbez2021revisiting}
\bibfield{author}{\bibinfo{person}{Richard Skarbez}, \bibinfo{person}{Missie Smith}, {and} \bibinfo{person}{Mary~C Whitton}.} \bibinfo{year}{2021}\natexlab{}.
\newblock \showarticletitle{Revisiting Milgram and Kishino's reality-virtuality continuum}.
\newblock \bibinfo{journal}{\emph{Frontiers in Virtual Reality}}  \bibinfo{volume}{2} (\bibinfo{year}{2021}), \bibinfo{pages}{647997}.
\newblock


\bibitem[Sutherland(1968)]%
        {sutherland68}
\bibfield{author}{\bibinfo{person}{I. Sutherland}.} \bibinfo{year}{1968}\natexlab{}.
\newblock \showarticletitle{A head-mounted three dimensional display}. In \bibinfo{booktitle}{\emph{Proc. Fall Joint Computer Conference}}. \bibinfo{publisher}{Thompson Books}, \bibinfo{address}{Wash. D.C.}, \bibinfo{pages}{757--764}.
\newblock


\bibitem[Teixeira(2010)]%
        {martycoopermobilephone}
\bibfield{author}{\bibinfo{person}{Tania Teixeira}.} \bibinfo{year}{23 April 2010}\natexlab{}.
\newblock \showarticletitle{Meet Marty Cooper - the inventor of the mobile phone}.
\newblock \bibinfo{journal}{\emph{BBC News}} (\bibinfo{year}{23 April 2010}).
\newblock


\bibitem[Thatcher(2017)]%
        {thatcher2017sousveillant}
\bibfield{author}{\bibinfo{person}{J. Thatcher}.} \bibinfo{year}{2017}\natexlab{}.
\newblock \showarticletitle{SOUSVEILLANT MEDIA}.
\newblock \bibinfo{journal}{\emph{Understanding spatial media}} (\bibinfo{year}{2017}), \bibinfo{pages}{56}.
\newblock


\bibitem[Vertegaal and Shell(2008)]%
        {vertegaal2008attentive}
\bibfield{author}{\bibinfo{person}{R. Vertegaal} {and} \bibinfo{person}{J. Shell}.} \bibinfo{year}{2008}\natexlab{}.
\newblock \showarticletitle{Surveillance and sousveillance of gaze-aware objects}.
\newblock \bibinfo{journal}{\emph{Social Science Information}} \bibinfo{volume}{47}, \bibinfo{number}{3} (\bibinfo{year}{2008}), \bibinfo{pages}{275--298}.
\newblock


\bibitem[Weber(2012)]%
        {weber2012surveillance}
\bibfield{author}{\bibinfo{person}{K. Weber}.} \bibinfo{year}{June 30, 2012}\natexlab{}.
\newblock \showarticletitle{Surveillance, Sousveillance, Equiveillance: Google Glasses}.
\newblock \bibinfo{journal}{\emph{Social Science Research Network, Research Network Working Paper, pp. 1-3 - http://tinyurl.com/6nh74jl}} (\bibinfo{year}{June 30, 2012}).
\newblock


\bibitem[Weston and Jacques(2009)]%
        {Weston2009Embracing}
\bibfield{author}{\bibinfo{person}{D Weston} {and} \bibinfo{person}{P Jacques}.} \bibinfo{year}{2009}\natexlab{}.
\newblock \showarticletitle{Embracing the `sousveillance state'}. In \bibinfo{booktitle}{\emph{Proc. Internat. Conf. on The Future of Ambient Intelligence and ICT for Security}}. \bibinfo{address}{Brussels}, \bibinfo{pages}{81}.
\newblock
\newblock
\shownote{ICTethics, FP7-230368}.


\bibitem[Wiener(2019)]%
        {wiener2019cybernetics}
\bibfield{author}{\bibinfo{person}{Norbert Wiener}.} \bibinfo{year}{2019}\natexlab{}.
\newblock \bibinfo{booktitle}{\emph{Cybernetics or Control and Communication in the Animal and the Machine}}.
\newblock \bibinfo{publisher}{MIT press}.
\newblock


\end{thebibliography}
\end{document}